\documentclass[journal]{IEEEtran}
\usepackage{caption} 
\usepackage{soul}
\usepackage{lipsum}
\usepackage{indentfirst}
\usepackage{booktabs}
\usepackage{enumerate}
\usepackage{tabu}
\usepackage{tabularx}
\usepackage{multirow}
\usepackage{environ}         
\usepackage{etoolbox}        
\usepackage[dvipsnames]{xcolor}

\usepackage{cite}

\usepackage{graphicx}
\usepackage{wrapfig}
\usepackage[labelformat=simple]{subcaption}

\usepackage{float}
\usepackage{eurosym}
\usepackage{textpos}
\usepackage{fancyhdr}
\usepackage[pscoord]{eso-pic}
\newcommand{\placetextbox}[3]{
  \setbox0=\hbox{#3}
  \AddToShipoutPictureFG*{
    \put(\LenToUnit{#1\paperwidth},\LenToUnit{#2\paperheight}){\vtop{{\null}\makebox[0pt][c]{#3}}}%
  }%
}%

\usepackage{amsfonts, eqnarray, amsmath, amssymb}
\usepackage{mathtools}
\usepackage{xfrac}
\usepackage{breqn}
\usepackage{bm}
\usepackage{resizegather}

\makeatother
\newtheorem{theorem}{Theorem}
\newtheorem{proposition}[theorem]{Proposition}

\usepackage{algorithm}
\usepackage{algpseudocode}
\usepackage{etoolbox}

\makeatletter
\newcommand*{\algrule}[1][\algorithmicindent]{%
  \hspace*{.2em}
  \vrule 
  \hspace*{\dimexpr#1-.2em-.4pt}%
}

\newcommand{\StatePar}[1]{%
  \State\parbox[t]{\dimexpr\linewidth-\ALG@thistlm}{\strut #1\strut}%
}
\renewcommand{\ALG@beginalgorithmic}{\offinterlineskip}

\newcount\ALG@printindent@tempcnta
\def\ALG@printindent{%
  \ifnum \theALG@nested > 0
    \ifx\ALG@text\ALG@x@notext
    \else
      \unskip
      \ALG@printindent@tempcnta=1
      \loop
        \algrule[\csname ALG@ind@\the\ALG@printindent@tempcnta\endcsname]%
        \advance \ALG@printindent@tempcnta 1
        \ifnum \ALG@printindent@tempcnta<\numexpr\theALG@nested+1\relax
      \repeat
        \fi
    \fi
}
\patchcmd{\ALG@doentity}{\noindent\hskip\ALG@tlm}{\ALG@printindent}{}{\errmessage{failed to patch}}
\makeatother

\algrenewcommand\algorithmicend{\strut\textbf{end}}
\algrenewcommand\algorithmicdo{\strut\textbf{do}}
\algrenewcommand\algorithmicwhile{\strut\textbf{while}}
\algrenewcommand\algorithmicfor{\strut\textbf{for}}
\algrenewcommand\algorithmicforall{\strut\textbf{for all}}
\algrenewcommand\algorithmicloop{\strut\textbf{loop}}
\algrenewcommand\algorithmicrepeat{\strut\textbf{repeat}}
\algrenewcommand\algorithmicuntil{\strut\textbf{until}}
\algrenewcommand\algorithmicprocedure{\strut\textbf{procedure}}
\algrenewcommand\algorithmicfunction{\strut\textbf{function}}
\algrenewcommand\algorithmicif{\strut\textbf{if}}
\algrenewcommand\algorithmicthen{\strut\textbf{then}}
\algrenewcommand\algorithmicelse{\strut\textbf{else}}

\algrenewcommand\algorithmicrequire{\strut\textbf{Input:}}
\algrenewcommand\algorithmicensure{\strut\textbf{Output:}}

\let\oldState\State
\renewcommand{\State}{\oldState\strut}

\usepackage{array}

\usepackage{url}

\NewEnviron{resizealign}{\sbox0{
    $\begin{matrix}\displaystyle\BODY\end{matrix}$}%
  \sbox1{$(\theequation)$}%
  \sbox2{\parbox{\dimexpr \wd0 + 2\wd1}%
    {\begin{align}\BODY\end{align}}}
  \noindent\resizebox{\columnwidth}{!}{\usebox2}%
}
    
\begin{document}
%
\title{Fairness Guaranteed and Auction-based x-haul and Cloud Resource Allocation in Multi-tenant O-RANs}
\author{Sourav~Mondal,~\IEEEmembership{Member,~IEEE}
        and~Marco~Ruffini,~\IEEEmembership{Senior~Member,~IEEE}%
\thanks{S. Mondal and M. Ruffini are with CONNECT Centre for Future Networks and Communication, Trinity College Dublin, University of Dublin, Dublin 2, Ireland (e-mail: somondal@tcd.ie, marco.ruffini@tcd.ie).}
\thanks{This work is financially supported by EU H2020 EDGE/MSCA (grant 713567) and Science Foundation Ireland (SFI) grants 17/CDA/4760 and 13/RC/2077\_P2.}} 

\placetextbox{0.5}{0.04}{This article is accepted for publication in IEEE Transactions on Communications. Copyright @ IEEE 2023.}%

\maketitle

\begin{abstract}
The open-radio access network (O-RAN) embraces cloudification and network function virtualization for base-band function processing by dis-aggregated radio units (RUs), distributed units (DUs), and centralized units (CUs). These enable the cloud-RAN vision in full, where multiple mobile network operators (MNOs) can install their proprietary or open RUs, but lease on-demand computational resources for DU-CU functions from commonly available open-clouds via open x-haul interfaces. In this paper, we propose and compare the performances of \emph{min-max fairness} and \emph{Vickrey-Clarke-Groves (VCG) auction}-based x-haul and DU-CU resource allocation mechanisms to create a multi-tenant O-RAN ecosystem that is sustainable for small, medium, and large MNOs. The min-max fair approach \emph{minimizes the maximum OPEX of RUs} through cost-sharing proportional to their demands, whereas the VCG auction-based approach \emph{minimizes the total OPEX for all resources utilized while extracting truthful demands from RUs}. We consider time-wavelength division multiplexed (TWDM) passive optical network (PON)-based x-haul interfaces where PON virtualization technique is used to flexibly provide optical connections among RUs and edge-clouds at macro-cell RU locations as well as open-clouds at the central office locations. Moreover, we design efficient heuristics that yield significantly better economic efficiency and network resource utilization than conventional greedy resource allocation algorithms and reinforcement learning-based algorithms.
\end{abstract}

\begin{IEEEkeywords}
Min-max fairness, multi-tenant Open-RAN, reinforcement learning, resource allocation, VCG auction.
\end{IEEEkeywords}

%
\IEEEpeerreviewmaketitle

\vspace{-0.5\baselineskip}
\section{Introduction} \label{sec1}
\begin{figure}[t!]
\centering
\includegraphics[width=\columnwidth]{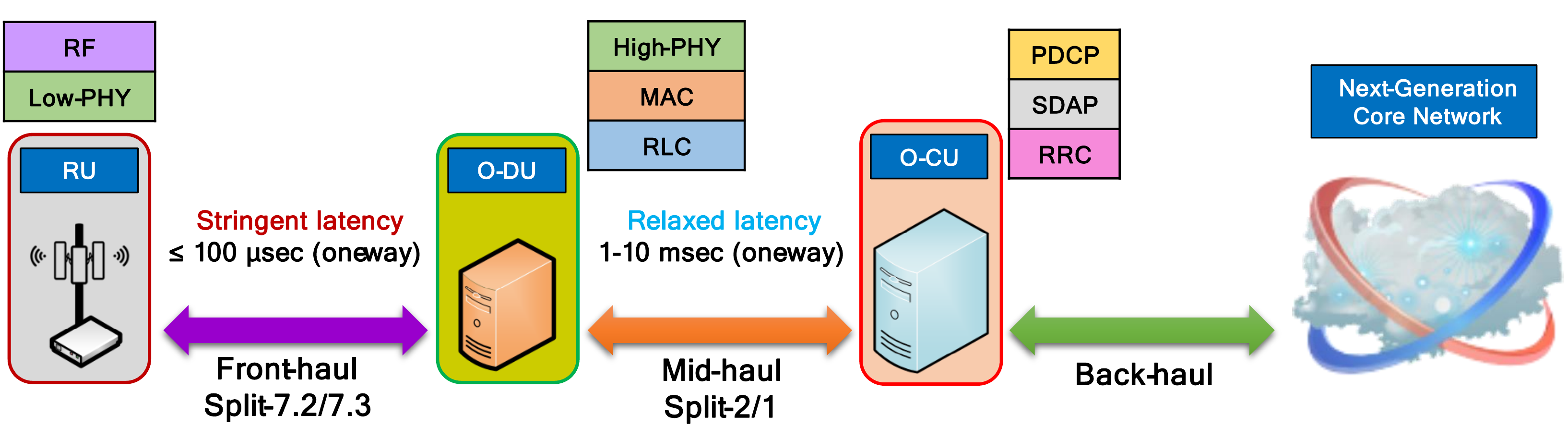}
\caption{A schematic diagram showing O-RAN architecture with functions of RU, O-DU, and O-CU and their corresponding interfaces.}
\label{o-ran}
\end{figure}

The fifth-generation (5G) radio access networks (RANs) are standardized to meet a diverse set of QoS requirements to support broadband, low-latency, and machine-type communications. Applications like mixed reality, telesurgery, high-definition video streaming, and Industrial Internet-of-Things, to name a few, will be free from the spectrum crunch and network resource scarcity issues of the legacy RANs. However, the existing mobile networks with their ``one size fits all" architecture lack sufficient flexibility and intelligence for efficient catering of such requirements \cite{6G_vision}. Therefore, the necessity for a major architectural revolution is envisaged for beyond 5G and sixth-generation (6G) RANs. Over the past few years, major mobile network operators (MNOs) across the globe are collaborating within the \emph{Open-RAN (O-RAN) Alliance} to standardize an open and smart RAN architecture that can perform complex RAN management with the aid of software-defined networking (SDN), network function virtualization (NFV), and edge computing (EC) technologies \cite{oran}. This architecture typically follows 3GPP recommendations where the RUs perform low-PHY functions (typically split 7.2 and 7.3), while high-PHY, MAC, RLC, RRC, and PDCP functions are processed by the DU-CUs that can be hosted on OLT-Clouds with commercial off-the-shelf (COTS) hardware, as shown in Fig. \ref{o-ran}. Recently, the IEEE P1914.1 standardization working group was created to specify the next-generation front-haul interface (NGFI). The RU-DU interface is known as the \emph{NGFI-I}, or the \emph{front-haul} (maximum one-way latency bound = 100 $\mu$sec), and the DU-CU interface is known as the \emph{NGFI-II} or the \emph{mid-haul} (maximum one-way latency bound = 1 msec) \cite{ngfi2}. The interface beyond CU to the 5G core is known as the \emph{back-haul}; hence, the general term \emph{x-haul} is used.\par
\hspace{-0.5\baselineskip}
\setlength{\textfloatsep}{5pt}
The incorporation of open clouds for DU-CU function processing over the open front/mid-haul interfaces in the O-RAN architecture creates new business opportunities for small, medium, and large MNOs as well as network service providers (NSPs) \cite{aceg1}. In turn, this creates a \emph{multi-tenant O-RAN ecosystem} where several MNOs deploy their RUs with macro and small-cell coverage over a certain geographic area but procure front/mid-haul and DU-CU function processing resources from the open and shared resource pool provided by various NSPs \cite{smsng}. The primary benefit of this multi-tenant O-RAN architecture is minimization of the CAPEX and OPEX for the MNOs. The techno-economic analysis in \cite{tech_eco} shows that $\sim$40\% CAPEX and $\sim$15\% OPEX over 5 years can be reduced by adopting SDN-based architectures for mobile network virtualization. In practice, government, municipality, or an alliance of MNOs can be the NSP that owns the open x-haul and cloud resources and distribute the resources among the MNOs. On the other hand, a competitive market model can also be created where the MNOs compete against each other or form opportunistic coalitions for procuring their required x-haul and cloud resources. These observations motivate us to propose efficient resource allocation mechanisms that create a multi-tenant O-RAN ecosystem that is sustainable for small, medium, and large MNOs.\par
%
The cloud servers installed at a central office (CO) or optical line terminal (OLT) locations are referred to as OLT-Clouds, but their significant intermediate distance may become disadvantageous for supporting low-latency applications and front-haul interfaces (typical PON length $\geq 10$ km). This hurdle can be overcome by installing Edge-Clouds at macro-cell RU locations to host DU-CU and \emph{local core} functions for some of the neighboring small-cell RUs \cite{pon_oran}. Moreover, efficiently utilizing geographically distributed Edge-Clouds can lead to a better cost efficiency of a RAN than centralized OLT-Clouds. Nonetheless, the RUs supporting latency-tolerant broadband applications can be connected to OLT-Cloud and 5G core without such issues. Therefore, we consider the TWDM-PON architecture proposed in \cite{sandip} as the x-haul interfaces to create a logical mesh topology that facilitates the small-cell RUs to be connected with OLT-Clouds at CO locations or Edge-Clouds at macro-cell RU locations in a flexible manner. This architecture \emph{supports East-West communication} along with traditional North-South communication and its efficiency over similar architectures in literature was also proven in \cite{sandip}.\par
We critically observe that a large body of the existing literature mainly focuses on allocating computational resources only and ignores communication resources of the x-haul interfaces. Moreover, while connecting RUs from different MNOs to either Edge-Cloud or OLT-Cloud over the open front/mid-haul interfaces, the OPEX of the RUs are calculated by either of the well-known methods like uniform sharing, utility maximization, min-max fairness, and proportional fairness \cite{fair}. Note that this resource allocation problem can be considered as an \emph{assignment problem}, but the RUs can not demand any specific amount of resources as in the conventional setting. Each RU only knows its front/mid-haul datarate and RU-DU-CU processing requirements corresponding to its split option. After all the RUs inform their respective front/mid-haul datarate to the NSP, sufficient resources are allocated by the NSP such that the front/mid-haul data generated by the RUs in each \emph{slot duration} (5G slot duration can be 125, 250, 500, or 1000 $\mu$sec) are transmitted and processed within the maximum latency bounds. Therefore, in the uniform sharing approach, when the cost of total utilized resources is uniformly distributed among the RUs, inefficiency may arise if RUs with lower resource requirements pay higher prices. In the utility maximization approach, the profit of the NSP is maximized while RUs are connected to Edge/OLT-Clouds. Hence, the RUs from wealthy MNOs will get priority and the RUs from poor MNOs may suffer from resource starvation at high-load conditions. The proportional fairness is a fair resource allocation method where fairness is achieved through maximization of a logarithmic utility function.\par 
Nevertheless, in this paper, we embrace the \emph{min-max fairness approach with proportional cost sharing} method, where we connect the RUs to Edge/OLT-Clouds such that the maximum OPEX of the RUs is minimized by allocating resources proportional to their demands and satisfy their latency requirements. Also, the RUs from different MNOs are fairly chosen for allocation such that poor MNOs do not suffer heavily during high-load conditions. We design this method for creating a multi-tenant O-RAN ecosystem where all the small, medium, and large MNOs get fair opportunities for OPEX minimization. However, the decisions made by this scheme strongly depend on the \emph{revealed resource demands of the RUs to the NSPs} and the RUs may not be always truthful in revealing their resource demands if there exist opportunities to gain extra incentives from the market. This motivates us to design a \emph{Vickrey-Clarke-Groves (VCG) auction-based mechanism} that allocates resources to RUs while minimizing the cost of total utilized resources but uses a special payment rule that enforces truthful revelation of resource requirements as a weakly dominant strategy equilibrium for the RUs \cite{nyto_auct}. Our contributions in this paper are:
\begin{enumerate}[(a)]
\item We propose a multi-tenant O-RAN architecture where RUs from small, medium, and large MNOs can be connected to Edge/OLT-Clouds for their DU-CU functions for low-latency and broadband applications in a sustainable manner over TWDM-PON-based front/mid-haul interfaces via East-West and North-South links.
\item We formulate an integer non-linear program (INLP) for the min-max fair resource allocation. In this formulation, we minimize the maximum cost for leasing front/mid-haul and DU-CU resources of each RU (resource allocation is proportional to demand) while satisfying the latency requirements of the low-latency and broadband applications.
\item We formulate a second INLP for the VCG auction-based resource allocation. In this formulation, we minimize the total cost for leasing front/mid-haul and DU-CU resources of all the RUs. Moreover, a payment rule is designed that ensures truthful revelation of resource demands of the RUs to prevent them from taking unfair advantages while paying for consumed resources.
\item We design polynomial-time algorithms for efficient implementation of the min-max fair and VCG auction formulations. Furthermore, we compare the economic efficiency and network resource utilization achieved by our proposed algorithms against state-of-the-art nearest-first (greedy) and reinforcement learning (RL)-based (multi-arm bandit) algorithms through numerical evaluation to showcase their usefulness in practice.
\end{enumerate}
\par The rest of this paper is organized as follows. Section \ref{sec2} reviews some related works. Section \ref{sec3} describes the multi-tenant O-RAN architecture. Section \ref{sec4} presents the system model. Section \ref{sec5} presents the min-max fairness, VCG auction, and baseline (greedy nearest-first and RL-based) methods. Section \ref{sec6} presents numerical evaluation results. Finally, Section \ref{sec7} provides the concluding remarks.

\begin{figure*}[t!]
\centering
\includegraphics[width=0.75\textwidth,keepaspectratio]{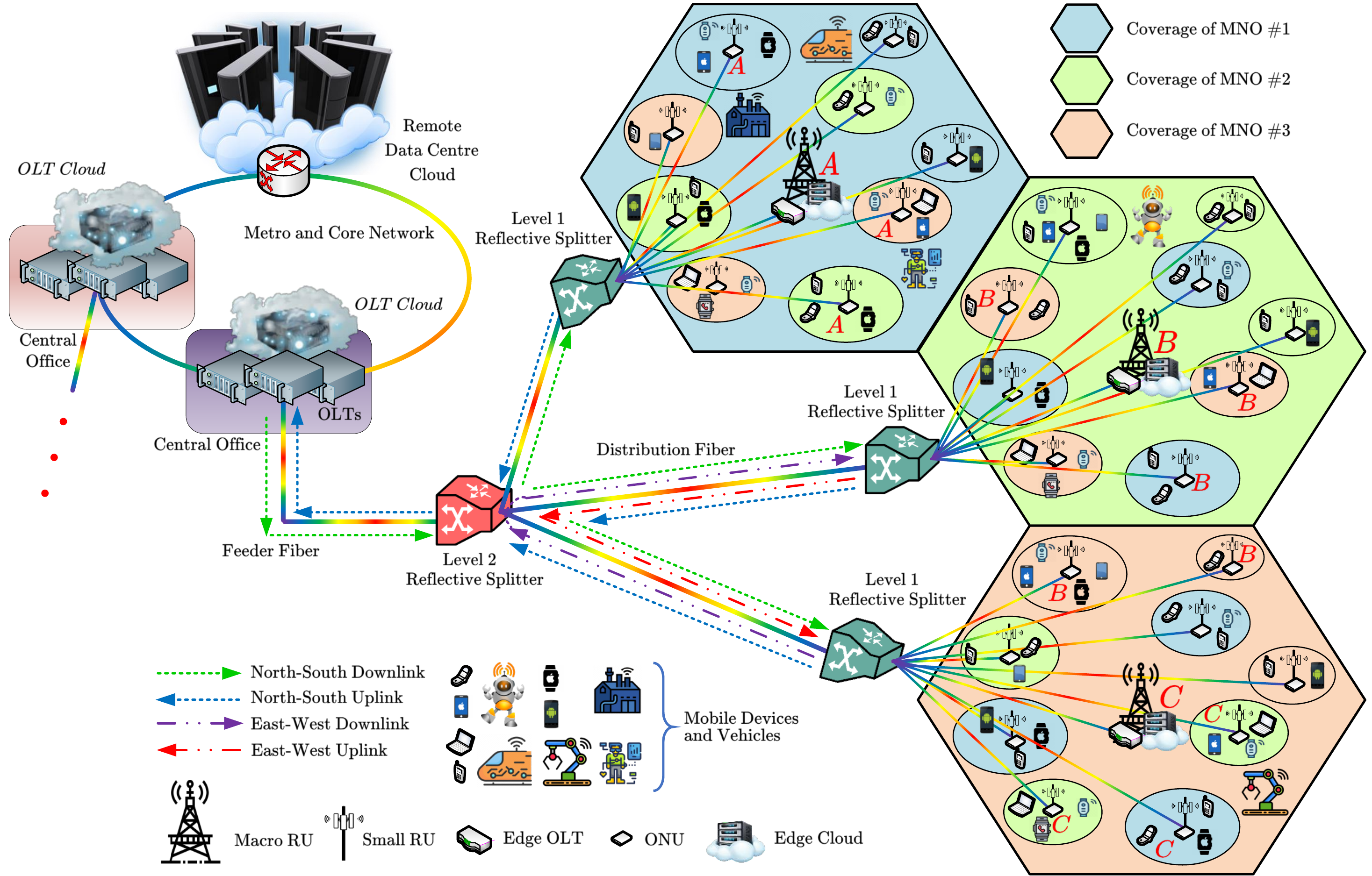}
\caption{The proposed TWDM-PON-based multi-tenant O-RAN architecture where RUs from multiple MNOs (hexagonal macro-cell and circular small-cell coverage area are shown in blue, green, and orange for three different MNOs) can be connected to Edge-Cloud or CO OLT-Cloud via the North-South or East-West virtual-PONs (indicated by red A, B, C) for the respective DUs and CUs.}
\label{architecture}
\end{figure*}
\setlength{\textfloatsep}{1pt}

\section{Review of Related Works} \label{sec2}
Resource allocation and management problems are fundamental research challenges in any networking environment and a large volume of literature exists on this area spanning across all types of network scenarios \cite{res_alloc_bk2}. The O-RAN for beyond 5G/6G mobile communication systems is no exception to this as its flexibility in terms of bandwidth, latency, and QoS requirements introduces several interesting research challenges \cite{oran_app}. Before the O-RAN architecture was proposed, several resource allocation or radio resource head (RRH) to base-band unit (BBU) assignment problems were solved using mathematical optimization and game theoretic tools for the Cloud-RAN (C-RAN) architecture by the authors of \cite{yao,cran_mgmt,cran_opt1,cran_opt2}. The authors of \cite{cran_ra} designed a dynamic two-stage mechanism for downlink resource allocation and BBU-RRH assignment in C-RAN. The authors of \cite{rrh-bbu1} investigated a joint RRH-BBU association and energy sharing problem to minimize brown energy usage. Again, the authors of \cite{rrh-bbu4} investigated the RRH-BBU mapping problem to minimize the network power consumption by reducing the number of active BBUs. Moreover, the authors of \cite{rru-bbu5} studied the joint RRH clustering and RRU activation problem with QoS constraints to minimize the energy consumption of RRHs. The authors of \cite{tenant1} demonstrated a multi-vendor multi-standard PON for 5G x-haul that performs the control and management operations by SDN/NFV technologies.\par
After the formation of the O-RAN Alliance, as the standardization of virtualized RAN started, researchers from academia and industry started to propose various interesting solutions to overcome O-RAN deployment and resource management challenges. Recently, the authors of \cite{oran_srvy} provided a very elaborate overview of the architecture and components of O-RAN, explored artificial intelligence (AI)-based use cases, and discussed various research opportunities across different engineering sectors. The authors of \cite{oran_snsr} provided detailed discussions on the ongoing O-RAN Alliance standardization activities with various analyses supported by a study of the traffic steering use case in a modular way following the open networking approach. We also shared several insights on optical transmission network (OTN) and optical distributed network (ODN)-based front/mid-haul network design for O-RANs from our observations in \cite{oran_sm_tnsm}. Alongside these, the authors of \cite{oran_tnsm} formulated a two-step mixed-integer programming problem for finding the optimal power allocation, physical resource block (PRB) assignment, the number of virtual network functions (VNFs), and the number of RUs. The authors of \cite{oran_ra} modeled the RU-DU assignment problem as a 2D bin packing problem and proposed a deep reinforcement learning-based self-play method to achieve efficient RU-DU resource management. Moreover, the authors of \cite{oran_team} designed a team learning algorithm for implementing a near-real-time (near-RT) radio intelligent controller (RIC) of O-RAN. However, neither of the aforementioned works focused on the challenges of designing flexible front/mid-haul interfaces between RUs and Edge/OLT-Clouds. Moreover, no comparative analysis is available between the conventional greedy heuristics and learning-based resource allocation algorithms.\par
Another important aspect of the O-RAN architecture, which essentially evolves from the C-RAN architecture, is its natural ability to facilitate multi-tenancy, i.e., a pool of network resources can be shared among multiple MNOs \cite{mlt_op_ran}. The multi-operator RAN (MORAN) allows two or more MNOs to share every component of a RAN except the radio carriers, whereas the multi-operator core network (MOCN) allows two or more core networks to share the same RAN or the carriers \cite{WirelessMoves}. In \cite{tenant2}, the authors demonstrated a virtual network controller enabled multi-tenant virtual network on top of multi-technology OTNs. We also performed some initial studies on the resource allocation problem for multi-tenant O-RAN ecosystems in \cite{sm_mnmx}. However, a more detailed investigation of system performance, economic analysis, and robust resource allocation mechanisms implementable in a practical competitive market scenario is required.
%

\section{Multi-tenant O-RAN Architecture} \label{sec3}
Fig. \ref{architecture} shows the considered O-RAN architecture in a multi-tenant scenario where multiple MNOs install neighboring RUs with hexagonal macro-cell and circular small-cell coverage. Each MNO pays a fee for leasing networking (i.e., for x-haul) and computing resources (i.e., DU-CU processing at Edge/OLT-Clouds) according to a certain payment scheme. Furthermore, all the MNOs need to pay a default price to the mediator, acting as the open platform provider to cover the cost of the resources required for RAN management and control plane operations. Recently, ITU-T has drafted recommendations for using TWDM-PONs as an optical front/mid-haul solution as TWDM-PONs can support 100 Gbps or more aggregated datarate (i.e., in upcoming standardization) which can be scaled further by combining additional wavelengths \cite{itu_5g}. Moreover, other recent work has addressed PON slicing isolation \cite{marco1} and compliance with service level agreement (SLAs) \cite{marco2}. Thus, TWDM-PON-based interfaces are used to connect both the macro and small-cell RUs to a CO with multi-level \emph{reflective splitters}. These splitters are designed so that they can be dynamically reconfigured to pass through or reflect back (i.e., towards the end points) the desired set of wavelengths (the concept is taken from \cite{sandip}). The wavelengths that are passed through, establish the North-South communication links (downlink: green, uplink: blue), whereas the reflected wavelengths establish the East-West communication links (downlink: purple, uplink: red). In terms of network hierarchy, each level-1 reflective splitter aggregates multiple RUs and each level-2 reflective splitter aggregates multiple level-1 reflective splitters and all their respective RUs for a cost-efficient deployment. A set of level-1 reflective splitters are used to connect RUs and Edge-Clouds directly, while multiple level-1 reflective splitters are connected to a level-2 reflective splitter to reach through other PON branches.\par
A local connectivity between small-cells and macro-cells via East-West communication links can be achieved by installing an Edge-OLT at the macro-cell, while the small-cells can host a simple ONU. Control signals via the North-South communication links can be sent to these Edge-OLTs at macro-cells and ONUs at small-cells to create virtual-PON instances that communicate via the East-West communication links. For example, three virtual-PON instances are shown in Fig. \ref{architecture} and the ONUs and Edge-OLTs belonging to the same virtual-PON are labeled as A, B, and C in red. This direct communication enables ultra-low latency and ultra-low jitter communications as the signals remain in the optical domain while reflected back at the splitter. Note that ONUs in virtual-PON instance A communicate only via level-1 reflective splitter. The same occurs for instance C. However, ONUs in virtual-PON instance B can communicate via both level-1 and level-2 reflective splitters (i.e., they extend across two PON branches). Both OLT and Edge-Clouds can host the DU-CU functions. Although the OLT-Clouds host the main 5G core, the Edge-Clouds can be used to host local 5G cores \cite{pon_oran}. The back-haul traffic can be routed to the remote data centers via metro and core networks.\par
\begin{figure}[t!]
\centering
\includegraphics[width=\columnwidth]{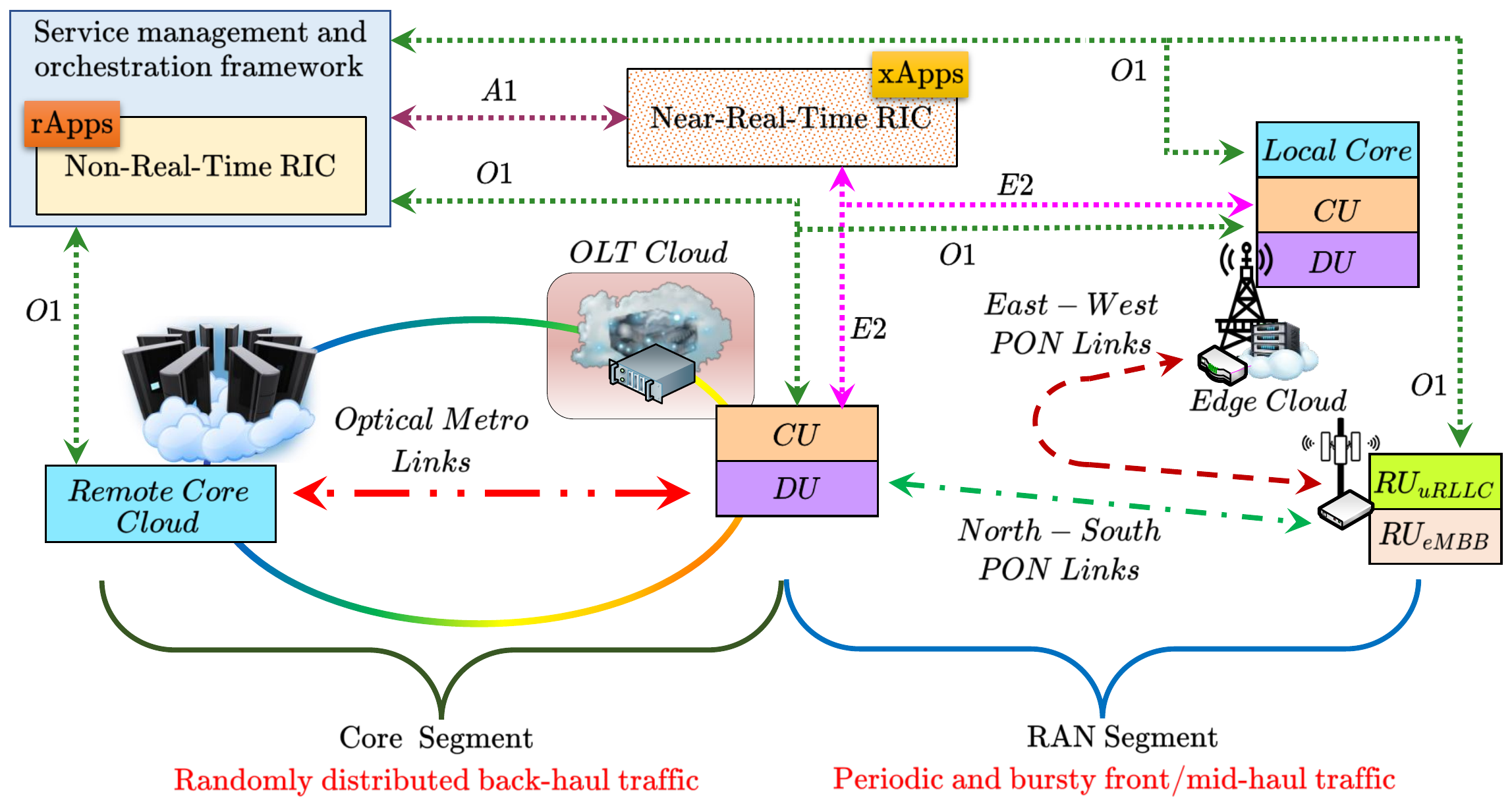}
\caption{A schematic diagram showing the user plane and control plane interfaces of the proposed TWDM-PON-based multi-tenant O-RAN architecture that supports uRLLC and eMBB services.}
\label{shared}
\end{figure}

Fig. \ref{shared} shows the user plane and control plane interfaces of the proposed TWDM-PON-based multi-tenant O-RAN architecture. The RUs for ultra-reliable and low-latency (uRLLC) services are prioritized to be connected to Edge-Clouds, whereas RUs for enhanced mobile broadband (eMBB) services can be flexibly connected to Edge/OLT-Clouds. Although Fig. \ref{o-ran} shows the most general schematic to highlight the flexible RAN deployment options provided by the O-RAN architecture, we choose to place DU and CU functions at a common Edge/OLT-Cloud because this is the most efficient configuration for uRLLC and eMBB applications in our judgment. The near-RT RIC mainly interacts with DUs and CUs through the $E2$ interface, whose control loops operate with a periodicity between 10 msec and 1 sec. The near-RT RIC consists of multiple applications called \emph{xApps} for per-UE controlled load-balancing, resource block management, interference detection and mitigation, QoS management, connectivity management, and seamless handover control \cite{colo}. Alongside this, the near-RT RIC is connected to the non-real time (non-RT) RIC by the $A1$ interface. This non-RT RIC is a component of the service management and orchestration (SMO) framework and consists of \emph{rApps} to complement the near-RT RIC for intelligent RAN operation and optimization on a time scale larger than 1 sec. Therefore, our proposed resource allocation algorithms in this paper can be implemented as control mechanisms that involve the periodic exchange of information and decision between RUs, Near-RT RIC, and Non-RT RIC. We consider that the RUs report their incoming resource demands to SMO every 1 sec over the $O1$ interface. Observing the information over a few seconds interval (operators decide based on the dynamicity of traffic), the rApps execute our proposed decision-making algorithms and pass on the decision to COs over the $O1$ interface. Accordingly, RUs are connected to Edge/OLT-Clouds over North-South or East-West TWDM-PON links. All the intermediate UE connectivity and handover management functions are handled by the xApps in near-RT RIC.
\setlength{\textfloatsep}{5pt}

\section{System Model} \label{sec4}
In this section, we describe the TWDM-PON-based front/mid-haul communication and RU-DU-CU function processing models considered for our problem formulation. The datarate for the front/mid-haul interface mainly depends on the split option chosen between RU and DU \cite{SCF}. With Split-7.2, all the radio frequency processing, fast Fourier transform (FFT)/inverse FFT, cyclic prefix removal/addition, digital beamforming, and resource element mapping are done at the RU. The datarate can be calculated as follows \cite{SCF}:
\begin{align}
    W_{7.2} = N_P \times N_{RB} \times N_{RB}^{SC} \times &N^{SF}_{sym} \times T^{-1}_{SF} \nonumber\\
    &\times \mu \times N_Q \times 2 \times \zeta, \label{eq01}
\end{align}
where, $N_P$ denotes the number of antenna ports, $N_{RB}$ denotes the number of resource blocks (RB), $N_{RB}^{SC}$ denotes the number of sub-carriers per RB, $T^{-1}_{SF}$ denotes sub-frame duration, $\mu$ denotes the maximum RB utilization, $N_Q$ denotes the quantizer bit resolution per I/Q dimension, and $\zeta$ denotes the front-haul overhead. With Split-7.3, precoding, layer mapping, and modulation are also done with the aforementioned tasks and the datarate is calculated as follows:
\begin{align}
    W_{7.3} = N_L &\times N_{RB} \times N^{RB}_{SC} \times N^{SF}_{sym} \times T^{-1}_{SF} \nonumber\\
    &\times \mu \times (1-\eta) \times N_Q \times \log_2(M_{mod}) \times \zeta, \label{eq02}
\end{align}
where, $N_L$ denotes the number of spatial layers, $\eta$ denotes resource overhead, and $M_{mod}$ denotes the modulation order. Note that 3GPP recommends Split-7.3 mainly to be used for downlink transmission, but Split-7.2 can be used for both uplink and downlink \cite{38.801}. As front/mid-haul data are transmitted as periodic bursts of Ethernet frames, the number of frames in a burst can be calculated as $\mathcal{B} = \lceil R_D\times \delta_t/\mathcal{P}\rceil$, where $R_D$ denotes the front/mid-haul datarate, $\delta_t$ denotes the burst interval duration, and $\mathcal{P}$ denotes the payload size of an Ethernet frame (1500 Bytes). Hence, the actual throughput of a flow can be calculated by $(\mathcal{B}\times F/\delta_t)$, where $F$ is the maximum Ethernet frame size (1542 Bytes). This data is transmitted over TWDM-PON and \emph{cooperative dynamic bandwidth allocation (Co-DBA)} protocol \cite{5g_fh_bw3} is used for coordinating RAN and PON capacity scheduling in the uplink transmission. Furthermore, it is crucial to note that sufficient communication resources should be available for each RU to transmit each burst of front/mid-haul data to DU-CU without failure to ensure a successful end-to-end communication.\par 
The total RU-DU-CU function processing effort per slot in Giga operations per second (GOPS) is given by \cite{mgain}:
\begin{align}
    \mathcal{C}_{RDC} = \left(3N_{a} + N_{a}^2 + \frac{1}{3} \times \mathcal{M}\times \Psi\times N_{L}\right)\times \frac{N_{RB}}{5}, \label{eq03}
\end{align}
\par \hspace{-1em}where, $N_{a}$ denotes the number of MIMO antennas, $\mathcal{M}$ denotes the number of modulation bits, and $\Psi$ denotes the coding rate. This total computational effort $\mathcal{C}_{RDC}$ is distributed among RU, DU, and CU based on the chosen intermediate split options. For example, 40\% processing is done by RU with Split-7.2, but 50\% processing is done by RU with Split-7.3. The remainder of the processing is done by the DU-CU and the total RU-DU-CU processing time can be computed by the polynomial expressions provided in \cite{bbu_lat}.

\section{Assignment of RUs to Edge/OLT-Clouds hosting DU/CU Functions} \label{sec5}
In this section, we formulate a min-max fairness-based and a VCG auction-based problem for connecting the RUs to some Edge/OLT-Cloud over front/mid-haul interfaces that host both the corresponding DU and CU functions. We also design some efficient heuristics for each of these problem formulations that can be implemented in practice. We consider that both macro-cell and small-cell RUs can support either or both uRLLC and eMBB applications. Thus, for mathematical convenience, we define two different sets for denoting RUs supporting uRLLC and eMBB applications, but some members of both these sets can be co-located and we need to optimally connect them to Edge-Clouds via East-West communication links or OLT-Clouds via North-South communication links. The min-max fairness-based resource allocation creates a multi-tenant O-RAN ecosystem where all the small, medium, and large MNOs get fair opportunities for OPEX minimization. In this scheme, each RU pays the price for allocated resources in proportion to their demand. However, to prevent affluent MNOs from influencing the fairness of resource allocation by revealing a higher resource demand, we formulate the VCG auction-based resource allocation problem with a different allocation and payment rule that makes each RU pay a price that is independent of their respective resource demand but dependent on the resource demands of other RUs. Thus, the RUs cannot gain any incentive by revealing any false resource demand. Although this formulation ensures truthful resource demand revelation from all the RUs, it cannot guarantee a fair OPEX for all MNOs because it minimizes the OPEX of the overall network and does not consider the OPEX of RUs individually. Therefore, NSPs can choose the min-max fairness resource allocation mechanism where truthful demand revelation is possible through strict market regulations (e.g., huge economic penalty or market ban on detection of false information). For an open and competitive market scenario, the NSPs can choose the VCG auction-based mechanism. Furthermore, we describe a nearest-first (greedy) and RL-based resource allocation mechanism for performance comparison.

\vspace{-\baselineskip}
\subsection{Min-Max Fairness Guaranteed Resource Allocation}
Our primary objective here is to allocate front/mid-haul and DU-CU resources for RUs such that the OPEX of RUs with worst/high values are minimized. We denote the set of uRLLC RUs by $\mathcal{R}_u = \{1,2,\dots,R_u\}$ and the set of eMBB RUs by $\mathcal{R}_m = \mathcal{R} \setminus \mathcal{R}_u$, where $\mathcal{R} = \{1,2,\dots,R_u,R_u+1,\dots,R_u+R_m\}$. Note that at each RU location, one RU for uRLLC services and one RU for eMBB services can coexist, whose data are scheduled to be transmitted at different PRBs within each slot. Also, we denote the set of Edge-Clouds by $\mathcal{E} = \{1,2,\dots,E\}$, and the set of OLT-Clouds by $\mathcal{Q} = \mathcal{Y}\setminus \mathcal{E}$, where $\mathcal{Y} = \{1,2,\dots,E,E+1,\dots,E+Q\}$. The \emph{binary variable} $x_{ry}$ denotes if an RU $r\in\mathcal{R}$ is connected to an Edge/OLT-Cloud $y\in\mathcal{Y}$, i.e.,
\begin{align*}
x_{ry}=\begin{cases}1; & \text{if RU $r$ and Edge/OLT-Cloud $y$ are connected} \\0; & \text{otherwise.}\end{cases}
\end{align*}
\par The parameter $z_{ry}$ indicates if RU $r\in\mathcal{R}$ and Edge/OLT-Cloud $y\in\mathcal{Y}$ can be connected over a virtual-PON (East-West or North-South) when its value is 1. The parameters $C_r$, $C_{\lambda}$, and $C_P$ denote the default cost to the mediator (\euro), the cost for throughput used (\euro/Gbps), and the cost for cloud resources leased (\euro/GOPS) by each RU $r$, respectively. As the Edge-Clouds are located at some macro-cell RU location, they can be owned by the respective MNO, and the attached RUs from the same MNO do not need to pay the costs of computational resources. The neutral NSP can also own the Edge-Clouds, but it needs to provide some price discount to the respective MNOs. To incorporate these facts, we incorporate a discount factor $\gamma_{ry}\in[0,1]$ where $\gamma_{ry}=0$ indicates full discount and $\gamma_{ry}=1$ indicates no discount. The parameters $W_r^{UL}$ and $W_r^{DL}$ denote the uplink and downlink front/mid-haul datarate of RU $r$. The parameters $B_y^{UL}$, $B_y^{DL}, \forall y\in\mathcal{E}$ denote the maximum uplink, and downlink throughput of the East-West TWDM-PON links and $B_y^{UL}$ and $B_y^{DL}, \forall y\in\mathcal{Q}$ denote the maximum uplink and downlink throughput of the North-South TWDM-PON links. The maximum throughput of each PON link can vary according to the number of configured wavelengths. The parameters $\eta_r^{UL}$ and $\eta_r^{DL}$ denote the required uplink and downlink GOPS/slot, $H_r^{UL}$ and $H_r^{DL}$ denote the available uplink and downlink GOPS/slot for RU processing. The parameters $\Gamma_r^{UL}$ and $\Gamma_r^{DL}$ denote the required GOPS/slot for DU-CU processing of RU $r$ and $G_y^{UL}$, $G_y^{DL}$ denote maximum available GOPS/slot at Edge/OLT-Clouds $y$. The parameter $\theta_{ry}$ denotes the burst interval over which data are transmitted from ONUs connected to RU $r$ in East-West or North-South TWDM-PONs. Finally, the parameters $\Delta_r^{H}$ and $\Delta_r^{RDC}$ denote the maximum one-way front/mid-haul latencies and total RU-DU-CU processing for RU $r$. Now, we formulate the min-max fair resource allocation problem for a multi-tenant O-RAN ecosystem as follows:
\begin{align}
    &\mathcal{P}_1:  \min_{x_{ry}}\max_r \left\{\sum_{y\in\mathcal{Y}} \left(C_r + C_\lambda B_{ry} + \gamma_{ry}C_P G_{ry}\right ) x_{ry} \right\} \label{eq04}\\
    & \text{subject to} \quad x_{ry} \leq z_{ry}, \forall r\in\mathcal{R}, y\in\mathcal{Y}, \label{eq05}\\
    & \hspace{4.7em} \sum\nolimits_{y\in\mathcal{Y}} x_{ry} \leq 1, \forall r\in\mathcal{R}, \label{eq06}
\end{align} \vspace{-\baselineskip}
\begin{align}
    & B_{ry} = \left(\frac{W_r^{UL} B_y^{UL}}{\varepsilon + \sum_r x_{ry} W_r^{UL}} \right)  + \left(\frac{W_r^{DL} B_y^{DL}}{\varepsilon + \sum_r x_{ry} W_r^{DL}} \right), \nonumber\\
    & \hspace{16.6em} \forall r\in\mathcal{R}, y\in\mathcal{Y}, \label{eq07} \\
    & G_{ry} = \left(\frac{\Gamma_r^{UL} G_y^{UL}}{\varepsilon + \sum_r x_{ry} \Gamma_r^{UL}}\right)  + \left(\frac{\Gamma_r^{DL} G_y^{DL}}{\varepsilon + \sum_r x_{ry} \Gamma_r^{DL}}\right), \nonumber\\
    & \hspace{16.6em} \forall r\in\mathcal{R}, y\in\mathcal{Y}, \label{eq08}\\
    & x_{ry}\left\{\delta_{ry} +\frac{D_{ry}}{v_l} \right\} + {\left\lceil \frac{\theta_{TTI}}{\theta_{ry}} \right\rceil} \left\{\frac{\sum_r x_{ry} W_r^{UL}\theta_{ry}}{B_y^{UL}} \right\} \leq \Delta_r^{H}, \nonumber\\
    & \hspace{16.6em} \forall r\in\mathcal{R}, y\in\mathcal{Y}, \label{eq09}\\
    & x_{ry}\left\{\frac{D_{ry}}{v_l} \right\} + {\left\lceil \frac{\theta_{TTI}}{\theta_{ry}} \right\rceil} \left\{\frac{\sum_r x_{ry} W_r^{DL}\theta_{ry}}{B_y^{DL}} \right\} \leq \Delta_r^{H}, \nonumber\\
    & \hspace{16.0em} \forall r\in\mathcal{R}, y\in\mathcal{Y}, \label{eq10} \\
    & \frac{\eta_r^{UL}}{H_r^{UL}} + \left\{\frac{\sum_r x_{ry} \Gamma_r^{UL}}{G_y^{UL}} \right\} \leq \frac{\Delta_r^{RDC}}{\theta_{TTI}}, \forall r\in\mathcal{R}, y\in\mathcal{Y}, \label{eq11} \\
    & \frac{\eta_r^{DL}}{H_r^{DL}} + \left\{\frac{\sum_r x_{ry} \Gamma_r^{DL}}{G_y^{DL}} \right\} \leq \frac{\Delta_r^{RDC}}{\theta_{TTI}}, \forall r\in\mathcal{R}, y\in\mathcal{Y}, \label{eq12}\\
    & \hspace{6.0em} x_{ry} \in \{0,1\}, \forall r\in\mathcal{R}, y\in\mathcal{Y}. \label{eq13}
\end{align}
\begin{table}[!t]
\centering
\caption{Network Parameters and Sets}
\label{table1}
\begin{tabular}{cl}
\toprule
\textbf{Symbol}      & \multicolumn{1}{c}{\textbf{Definition}}                                    \\ \toprule
$\mathcal{R}_u$      & Set of RUs for uRLLC services                                             \\ \hline
$\mathcal{R}_m$      & Set of RUs for eMBB services                                              \\ \hline
$\mathcal{R}$        & Set of all RUs present in the system ($\mathcal{R} = \mathcal{R}_u \cup \mathcal{R}_m$)                           \\ \hline
$\mathcal{E}$        & Set of Edge-Cloud locations           \\ \hline
$\mathcal{Q}$        & Set of OLT-Cloud locations            \\ \hline
$\mathcal{Y}$        & Set of all Edge/OLT-Cloud locations ($\mathcal{Y} = \mathcal{E} \cup \mathcal{Q}$)                        \\ \hline
$D_{ry}$             & Distance between RU $r\in\mathcal{R}$ and Edge/OLT-Cloud $y\in\mathcal{Y}$\\ \hline
$C_r$                & Default cost of participation to the mediator (\euro) \\ \hline
$C_{\lambda}$        & Cost for throughput used (\euro/Gbps) \\ \hline
$C_P$                & Cost for cloud resources leased (\euro/GOPS) \\ \hline
$W_r^{UL/DL}$        & The uplink and downlink front/mid-haul datarate of RU $r$ \\ \hline
$\Gamma_r^{UL/DL}$   & Required GOPS/slot for DU-CU processing of RU $r$ \\ \hline
$B_y^{UL/DL}$        & Maximum throughput of the TWDM-PON links \\ \hline
$G_y^{UL/DL}$        & Maximum available GOPS/slot at Edge/OLT-Clouds $y$ \\ \hline
$\eta_r^{UL/DL}$     & Required uplink and downlink GOPS/slot for RU processing \\ \hline
$H_r^{UL/DL}$        & Available uplink and downlink GOPS/slot for RU processing \\ \hline
$\Delta_r^H$         & Maximum one-way front/mid-haul latency of RU $r$         \\ \hline
$\Delta_r^{RDC}$     & Maximum RU-DU-CU processing latency of RU $r$  \\ \hline
$\theta_{slot}$       & Transmit time slot of RU $r$  \\ \hline
$\theta_{ry}$        & Transmission burst interval of TWDM-PON uplinks \\ \hline
$\delta_{ry}$        & The reduced waiting time for uplink data at ONUs \\ \hline
$v_l$                & Speed of light in optical fiber ($2\times 10^8$ m/s)                        \\ \bottomrule
\end{tabular}
\end{table}
\setlength{\textfloatsep}{5pt}
\vspace{-\baselineskip}
\par The objective function of the problem $\mathcal{P}_1$ is given by (\ref{eq04}), which indicates the minimization of maximum OPEX of each RU $r$. The first term is the default cost, the second term is the front/mid-haul throughput leasing cost, and the third term is the DU-CU function processing resources leasing cost. Note that \emph{the price for throughput and computational resources paid by each RU is proportional to their demands}. The constraint (\ref{eq05}) ensures that RU $r$ can be associated with Edge/OLT-Cloud $y$ only when an East-West or North-South connection exists and the constraint (\ref{eq06}) restricts RU $r$ to be connected to one Edge-Cloud or OLT-Cloud $y$ at most. The constraints (\ref{eq07}) indicates the allocated share of throughput to RU $r$ over front/mid-haul interface to Edge/OLT-Cloud $y$. Similarly, the constraint (\ref{eq08}) indicates the allocated share of GOPS to RU $r$ for DU-CU processing at Edge/OLT-Cloud $y$. Note that a very small constant $\varepsilon \approx 0$ is added to the denominator of each of the terms in (\ref{eq07})-(\ref{eq08}) to avoid division by zero. Furthermore, the constraint (\ref{eq09}) ensures that the uplink front/mid-haul latency from RU $r$ to Edge/OLT-Cloud $y$ is within $\Delta_r^{H}$. The parameter $\delta_{ry}$ denotes the \emph{average queuing latency} of uplink data due (considering the use of the Co-DBA mechanism). The second term with $x_{ry}$ indicates \emph{propagation latency} where the parameter $D_{ry}$ denotes the distance from RU $r$ to Edge/OLT-Cloud $y$ and $v_l$ denotes speed of light within fiber ($2\times10^5$ km/s). The third term indicates \emph{data transmission latency} where data is transmitted in multiple bursts of duration $\theta_{ry}$ within each TTI, $\theta_{TTI}$. Similarly, constraint (\ref{eq10}) ensures that the downlink front/mid-haul latency from RU $r$ to Edge/OLT-Cloud $y$ is within $\Delta_r^{H}$. Finally, the constraints (\ref{eq11})-(\ref{eq12}) ensure the uplink and downlink RU-DU-CU processing latencies are within $\Delta_r^{RDC}$, respectively. The first term indicates the RU processing latency and the second and third terms indicate the DU-CU processing latencies at Edge/OLT-Cloud $y$, respectively.

\vspace{-\baselineskip}
\subsection{Heuristic for the Min-Max Fair Resource Allocation}
We observe that $\mathcal{P}_1$ is an NP-hard problem and the primary reason behind the NP-hardness is that the locations of active Edge/OLT-Clouds are not known when we start to connect RUs to Edge/OLT-Clouds. Note that the problem $\mathcal{P}_1$ has a unique structure that converts a multi-objective problem into a single-objective problem such that standard optimization methods can be employed. In this case, we convert the minimization problem of OPEXs of multiple RUs into a minimization problem of the maximum OPEX of RUs. However, the problem $\mathcal{P}_1$ is still very inconvenient to solve due to the presence of $\max\{.,.\}$ function in the objective. Therefore, we need to reformulate this problem into an equivalent \emph{epigraph form} as follows:
\begin{align}
    &\mathcal{P}_1^r: \quad\quad\quad \min_{x_{ry},M} \quad M \label{eq14}\\
    &\text{subject to} \quad M \geq \sum\nolimits_{y\in\mathcal{Y}} \left(C_r + C_\lambda B_{ry} + \gamma_{ry} C_P G_{ry}\right ) x_{ry}, \nonumber\\
    & \hspace{19em} \forall r\in\mathcal{R}, \label{eq15}\\
    & \hspace{5em} \text{constraints } (5)-(13). \nonumber
\end{align}
\par It is straightforward to show that an optimal solution for $\mathcal{P}_1^r$ is also a solution for $\mathcal{P}_1$ \cite{moo}. Nonetheless, as both these problems are INLP, the evaluation of an optimal solution cannot be guaranteed in polynomial time. Hence, a heuristic algorithm is required. In general, we understand that OPEX of each RU in (\ref{eq04}) can be minimized if each front/mid-haul link and Edge/OLT-Cloud resources are leased by a maximum number of RUs while satisfying constraints (\ref{eq09})-(\ref{eq12}). In addition, we observe the following interesting property of optimal solutions of $\mathcal{P}_1^r$.
\begin{proposition} \label{prop1}
    \textit{An optimal solution of $\mathcal{P}_1^r$ can guarantee fairness if and only if the OPEX of an RU with lower resource requirements does not exceed the OPEX of an RU with higher resource requirements.}
\end{proposition}
%
\begin{algorithm}[t!]
\caption{Algorithm for min-max fair resource allocation} \label{alg1}
\hspace*{\algorithmicindent} \textbf{Input:} $\mathcal{R}, \mathcal{E}, \mathcal{Y}, D_{ry}, B_y^{U/DL}, W_r^{U/DL}, \Gamma_r^{U/DL}, G_r^{U/DL}$\\
\hspace*{\algorithmicindent} \textbf{Output:} Near-optimal solution: $x_{ry}^*$ and $\mathcal{C}_r^*$\\
\hspace*{\algorithmicindent} \textbf{Initialize:} Sort the elements of $\mathcal{R}$ in the increasing order\\
\hspace*{\algorithmicindent} of ($\max\{W_r^{DL},W_r^{UL}\}$) and/or ($\max\{\Gamma_r^{DL},\Gamma_r^{UL}\}$) while\\
\hspace*{\algorithmicindent} maintaining an uniform distribution of the percentage of\\ 
\hspace*{\algorithmicindent} ownership of MNOs;
\begin{algorithmic}[1]
\For{$r \leftarrow 1$ \textbf{to} $|\mathcal{R}|$}
    \If{$r = 1$} \Comment{\textit{choose best possible $r = 1$}}
        \State Set $\bar{\mathcal{Y}} \gets \mathcal{Y}$;
        \State Set $assign \gets 0$;
        \While{$assign \neq 1$ \textbf{and} $\bar{\mathcal{Y}} \neq \emptyset$}
            \State Find $y' = \arg\min_y \{D_{ry}\}, y'\in \bar{\mathcal{Y}}$;
            \If{constraints (\ref{eq05}), (\ref{eq09})-(\ref{eq12}) are satisfied}
                \State Set $x_{ry'}\gets 1$;
                \State Set $assign \gets 1$;
                \State Calculate $B_{ry}$, $G_{ry}$, and $\mathcal{C}_r$;
            \Else
                \State Set $\bar{\mathcal{Y}} \gets \bar{\mathcal{Y}}\setminus\{y'\}$;
            \EndIf
        \EndWhile
        \If{$assign \neq 1$ \textbf{and} $\bar{\mathcal{Y}} = \emptyset$}
            \State break; \Comment{\textit{infeasibility condition}}
        \EndIf
    \ElsIf{$1 < r \leq |\mathcal{R}|$}
         \State Find all $y\in \mathcal{Y}$ such that constraints (\ref{eq05}), (\ref{eq09})-(\ref{eq12})
         \State are satisfied for the current RU $r$ and create $\bar{\mathcal{Y}}$;
         \If{$|\bar{\mathcal{Y}}| \geq 1$}  \Comment{\textit{if at least one such $y$ exists}}
            \State Calculate all dummy OPEX values for RU $r$, 
            \State $\mathcal{C}_{ry}$, if $r$ was connected to each of $y\in \bar{\mathcal{Y}}$;
            \State Find $y' = \arg\min_y \{\mathcal{C}_{ry}\}, y'\in \bar{\mathcal{Y}}$;
            \State Set $x_{ry'}\gets 1$;
            \State Update $B_{ry}$, $G_{ry}$, and $\mathcal{C}_r, \forall r$ with $\sum_y x_{ry} = 1$;
        \Else
            \State Set $B_{ry} = 0$, $G_{ry} = 0$, and $\mathcal{C}_r = 0$;
         \EndIf
    \EndIf
\EndFor
\State \textbf{return} $x_{ry}$ and $\mathcal{C}_r$;
\end{algorithmic}
\end{algorithm}
\setlength{\textfloatsep}{5pt}
\par Please refer to Appendix A for the proof. In general, we can achieve the best possible value of $M$ if full-mesh connectivity is available among RU and Edge/OLT-Clouds. However, in practice, mostly partial-mesh connectivity can be observed, i.e., constraint (\ref{eq06}) along with constraints (\ref{eq09})-(\ref{eq12}) will have a strong influence on the solution. Nonetheless, in general, we are able to connect a higher number of RUs to Edge/OLT-Clouds if we start with lower resource requirements. Based on the above insights, we design a heuristic algorithm, summarized as Algorithm \ref{alg1}. At first, we sort the RUs in $\mathcal{R}$ in the increasing order of ($\max\{W_r^{DL},W_r^{UL}\}$) and ($\max\{\Gamma_r^{DL},\Gamma_r^{UL}\}$). This step is crucial to maintain consistency with Proposition \ref{prop1}. We also order the RUs such that the percentage of ownership of MNOs is uniformly maintained. Then we start to iteratively connect each RU $r$ to an Edge/OLT-Cloud $y$. For $r=1$, we initialize the flag $assign \gets 0$, the dummy set $\bar{\mathcal{Y}} \gets \mathcal{Y}$, and find the nearest $y' = \arg\min_y \{D_{ry}\}, y'\in \bar{\mathcal{Y}}$. If the constraints (\ref{eq05}), (\ref{eq09})-(\ref{eq12}) are satisfied for this $y'$, we set $x_{ry'}\gets 1$, $assign \gets 1$, and calculate the corresponding OPEX value $\mathcal{C}_r = \sum_y (C_r + C_\lambda B_{ry} + \gamma_{ry} C_P G_{ry} ) x_{ry}$. If this is not successful, then we remove $y'$ from $\bar{\mathcal{Y}}$ and continue this process until $\bar{\mathcal{Y}} = \emptyset$. In the subsequent iterations, i.e., for $1 < r \leq |\mathcal{R}|$, we find all $y\in \mathcal{Y}$ that satisfy constraints (\ref{eq05}), (\ref{eq09})-(\ref{eq12}) for the current RU $r$ and reinitialize the dummy set $\bar{\mathcal{Y}}$. If at least one such $y$ exists, i.e., $|\bar{\mathcal{Y}}| \geq 1$, then we calculate the dummy OPEX values $\mathcal{C}_{ry} = (C_r + C_\lambda B_{ry} + \gamma_{ry} C_P G_{ry})$ if RU $r$ was connected to each of the Edge/OLT-Cloud $y$. Then we find $y' = \arg\min_y \{\mathcal{C}_{ry}\}, y'\in \bar{\mathcal{Y}}$, set $x_{ry'}\gets 1$, and calculate the updated values of $B_{ry}$, $G_{ry}$, and $\mathcal{C}_r$ for all RUs. The first for loop iterates for $|\mathcal{R}|$ times to return $x_{ry}$ and $\mathcal{C}_r$ while finding a suitable Edge/OLT-Cloud $y$ from a set of maximum size $|\mathcal{Y}|$ at every iteration. Therefore, the worst-case time-complexity of this loop, as well as Algorithm \ref{alg1}, is $\mathcal{O}(|\mathcal{R}| \times |\mathcal{Y}|)$. Now, if we denote the optimal number of Edge/OLT-Cloud as $Y^*$, then the number of remaining RUs yet to be connected at every iteration $t$ of Algorithm \ref{alg1} is at most $R_t \leq R_{t-1}(1-\sfrac{1}{Y^*})$. Thus, at the end of all $|\mathcal{R}|$ iterations, we must have $1 \leq |\mathcal{R}|\times(1-\sfrac{1}{Y^*})^{Y_{|\mathcal{R}|}} = |\mathcal{R}|\times(1-\sfrac{1}{Y^*})^{Y^* \times \frac{Y_{|\mathcal{R}|}}{Y^*}}$. By using the Taylor series approximation $(1-\sfrac{1}{x})^{x} \approx \sfrac{1}{e}$, we get $1 \leq |\mathcal{R}|\times(\sfrac{1}{e})^{\frac{Y_{|\mathcal{R}|}}{Y^*}}$, or $Y_{|\mathcal{R}|} \leq Y^* \log_e(|\mathcal{R}|)$, where $Y_{|\mathcal{R}|}$ denotes the number of active Edge/OLT-Clouds after all $|\mathcal{R}|$ iterations. Hence, the solution produced by Algorithm \ref{alg1} approximates the optimal solution by a factor of $\mathcal{O}(\log_e(|\mathcal{R}|))$.

\vspace{-0.5\baselineskip}
\subsection{VCG Auction-based Resource Allocation}
Although the solution obtained from the min-max fairness method described in previous sub-sections is very efficient, the solution is dependent on the \emph{private information} like $W_r^{UL}$, $W_r^{DL}$, $\Gamma_r^{UL}$, and $\Gamma_r^{DL}$ shared by the RUs to the NSP. To prevent the RUs from sharing false information and gaining unfair incentives from the market, we design a VCG auction-based resource allocation method in this sub-section. Note that this problem is very close to an auction of multiple divisible items \cite{mult_auc} but with specific unique characteristics. Each RU wants to connect to an Edge/OLT-Cloud via some East-West or North-South TWDM-PON link to avail sufficient throughput and Edge/OLT-Cloud resources to successfully transmit its front/mid-haul data generated within each slot duration, satisfying the maximum latency bounds. Therefore, the private valuation function of each RU $r$ is given as:
\begin{align*}
V_r(x_{ry})=\begin{cases} (C_{\lambda} B_y + C_P G_y)x_{ry}; & \text{if $r$ is connected to}\\
 & \text{some $y$ with successful}\\
 & \text{front/mid-haul data}\\
 & \text{transmission,}
\\0; & \text{otherwise,}\end{cases}
\end{align*}
where, $B_y = (B_y^{UL}+B_y^{DL})$ and $G_y = (G_y^{UL}+G_y^{DL})$. Note that if an RU gets connected to some Edge/OLT-Cloud but fails to transmit its front/mid-haul data, then also its valuation of resources is zero. We assume that the NSP wants a fair market competition and hence, keeps the cost parameters $C_r$, $C_{\lambda}$, and $C_P$ the same for all the competing MNOs. At first, each RU $r$ submits their front/mid-haul datarate and DU-CU processing requirements to the NSP as a message $b_r = (\hat{W}_r^{UL}, \hat{W}_r^{DL}, \hat{\Gamma}_r^{UL}, \hat{\Gamma}_r^{DL})$. After receiving messages $\bm{b} = (b_1,b_2,\dots,b_{|\mathcal{R}|})$ from all RUs, the NSP solves the following cost-minimization problem while connecting a maximum number of RUs to Edge/OLT-Clouds and the solution $\hat{x}_{ry}^*$ can be considered as the \emph{allocation rule} \cite{Narahari}.
\begin{align}
    \mathcal{P}_2: \quad\quad\quad & \min_{x_{ry},t_y} \left\{\sum_{y\in\mathcal{Y}} \left(C_\lambda B_y + C_P G_y\right ) t_y \right\} \label{eq16}\\
    \text{subject to} \quad & t_y \leq x_{ry}, \forall r\in\mathcal{R}, y\in\mathcal{Y}, \label{eq17}\\
    & \text{constraints } (5)-(6), (9)-(13), \nonumber
\end{align}
where $t_y$ is a binary decision variable and the constraint (\ref{eq17}) implies that $t_y$ is equal to 1 if at least one RU $r$ is connected to the Edge/OLT-Cloud $y$.
\begin{proposition} \label{prop2}
    \textit{The allocation rule $\hat{x}_{ry}^*$ derived as a solution of $\mathcal{P}_2$ is allocatively efficient.}
\end{proposition}
\par Please refer to Appendix B for the proof. Although several other dominant-strategy truthful mechanisms exist in the \emph{quasi-linear setting}\footnote{The utility of agent $r$ with private valuation $v_r$ from obtaining a fraction $x$ of a divisible good at a price $p$ is $u_r(x,p) = v_r(x)-p$.}, only with Groves mechanisms, we can implement an allocative efficiency in dominant strategies among agents with arbitrary quasi-linear utilities. Observe that the valuation for resources of each RU $r$ is non-zero only if it is connected to some Edge/OLT-Cloud $y$ and gets sufficient resources to transmit its front/mid-haul data generated in each slot within the maximum latency bounds. Otherwise, if an RU $r$ is unallocated or fails to transmit its front/mid-haul data generated in each slot within the maximum latency bounds, its valuation is zero. Note that the same resources $B_y$ and $G_y$ are shared by multiple RUs that are allocated to Edge/OLT-Cloud $y$. Therefore, the net worth of resources allocated to each RU $r$, $\hat{\mathcal{C}}_r$ varies within $[0,(C_\lambda B_y + C_P G_y)]$. This implies that if only one RU is connected to an Edge/OLT-Cloud $y$, then it must bear the cost of the consumed resources $\hat{\mathcal{C}}_r = (C_\lambda B_y + C_P G_y)$ all alone. However, if multiple RUs are connected to an Edge/OLT-Cloud $y$, then the total cost is uniformly divided among them, i.e., $\hat{\mathcal{C}}_r = [(C_\lambda B_y + C_P G_y)/\sum_r \hat{x}_{ry}^*]$. Using this observation, we design a \emph{payment rule} for the RUs similar to Clarke's mechanism \cite{Narahari} as follows:
\begin{align}
    P_r(x_{ry},\bm{b}) = \left[\sum_{j\neq r} \tilde{\mathcal{C}}_j(x_{jy}^{-r*}(\bm{b}^{-r}),b_j) \right] - \left[\sum_{j\neq r} \tilde{\mathcal{C}}_j(x_{jy}^*(\bm{b}),b_j) \right] \nonumber
\end{align} \vspace{-0.5\baselineskip}
\begin{align}
    = \sum_{j\neq r} \left(\frac{C_\lambda B_y + C_P G_y}{\sum_{k\neq r} x_{ky}^*}\right) - \sum_{j\neq r} \left(\frac{C_\lambda B_y + C_P G_y}{\sum_k x_{ky}^*}\right), \label{eq18}
\end{align}
which can be interpreted as the total cost of all RUs other than $r$ under an efficient allocation when RU $r$ is absent in the system minus the total cost of all RUs other than $r$ under an efficient allocation when RU $r$ is present in the system. Note that $\hat{x}_{jy}^{-r*}$ or $\hat{x}_{ky}^{-r*}$ denote allocation rules when RU $r$ is absent in the system.
\setcounter{theorem}{0}
\begin{theorem}
    \textit{The payment rule (\ref{eq18}) ensures that truthful private information sharing to NSPs is a weakly dominant strategy for all RUs.}
\end{theorem}
\par Please refer to Appendix C for the proof. This also shows that the mechanism is \emph{weakly budget balanced} as $\sum_r P_r(\hat{x}_{ry}^*,\bm{b}) \geq 0$. With the aforementioned allocation and payment rules, the utility of each RU $r$ is defined as follows:
\begin{align}
    \mathcal{U}_r(\hat{x}_{ry}^*,\bm{b}) = V_r(\hat{x}_{ry}^*,\bm{b}) - P_r(\hat{x}_{ry}^*,\bm{b}). \label{eq19} 
\end{align}
\par Therefore, the \emph{individual rationality} of the RUs is always maintained in this mechanism as $\mathcal{U}_r(\hat{x}_{ry}^*,\bm{b}) \geq 0$. In addition to $P_r(\hat{x}_{ry}^*,\bm{b})$, each RU $r$ also needs to pay the default cost $C_r$ to the NSP if it is connected to some Edge/OLT-Cloud $y$.
\begin{algorithm}[t!]
\caption{Algorithm for VCG auction-based allocation} \label{alg2}
\hspace*{\algorithmicindent} \textbf{Input:} $\mathcal{R}, \mathcal{E}, \mathcal{Y}, D_{ry}, B_y^{U/DL}, W_r^{U/DL}, \Gamma_r^{U/DL}, G_r^{U/DL}$\\
\hspace*{\algorithmicindent} \textbf{Output:} Near-optimal solution: $x_{ry}^*$ and $\mathcal{C}_r^*$\\
\hspace*{\algorithmicindent} \textbf{Initialize:} Sort the elements of $\mathcal{R}$ in the increasing order\\
\hspace*{\algorithmicindent} of ($\max\{W_r^{DL},W_r^{UL}\}$) and ($\max\{\Gamma_r^{DL},\Gamma_r^{UL}\}$) while\\
\hspace*{\algorithmicindent} maintaining an uniform distribution of the percentage of\\ 
\hspace*{\algorithmicindent} ownership of MNOs;
\begin{algorithmic}[1]
\For{$r \leftarrow 1$ \textbf{to} $|\mathcal{R}|$}  \label{ln1}
    \If{$r = 1$} \Comment{\textit{choose best possible $r = 1$}}
        \State Set $\bar{\mathcal{Y}} \gets \mathcal{Y}$;
        \State Set $assign \gets 0$;
        \While{$assign \neq 1$ \textbf{and} $\bar{\mathcal{Y}} \neq \emptyset$}
            \State Find $y' = \arg\min_y \{D_{ry}\}, y'\in \bar{\mathcal{Y}}$;
            \If{constraints (\ref{eq05}), (\ref{eq09})-(\ref{eq12}) are satisfied}
                \State Set $x_{ry'}\gets 1$ and $t_{y'}\gets 1$;
                \State Set $assign \gets 1$;
                \State Calculate $\mathcal{C}_{tot}$ and $\tilde{\mathcal{C}}_{r}$;
            \Else
                \State Set $\bar{\mathcal{Y}} \gets \bar{\mathcal{Y}}\setminus\{y'\}$;
            \EndIf
        \EndWhile
        \If{$assign \neq 1$ \textbf{and} $\bar{\mathcal{Y}} = \emptyset$}
            \State break; \Comment{\textit{infeasibility condition}}
        \EndIf
    \ElsIf{$1 < r \leq |\mathcal{R}|$}
         \State Find all $y\in \mathcal{Y}$ such that constraints (\ref{eq05}), (\ref{eq09})-(\ref{eq12})
         \State are satisfied for the current RU $r$ and create $\bar{\mathcal{Y}}$;
         \If{$|\bar{\mathcal{Y}}| \geq 1$}  \Comment{\textit{if at least one such $y$ exists}}
            \State Calculate all dummy cost values for RU $r$, 
            \State $\mathcal{C}_{tot,y}$, if $r$ was connected to each of $y\in \bar{\mathcal{Y}}$;
            \State Find $y' = \arg\min_y \{\mathcal{C}_{tot,y}\}, y'\in \bar{\mathcal{Y}}$;
            \State Set $x_{ry'}\gets 1$ and $t_{y'}\gets 1$;
            \State Update $\mathcal{C}_{tot}$ and $\tilde{\mathcal{C}}_r, \forall r$ with $\sum_y x_{ry} = 1$;
        \Else
            \State Set $\tilde{\mathcal{C}}_r = 0$ and keep $\mathcal{C}_{tot}$ unchanged;
         \EndIf
    \EndIf
\EndFor             \label{ln31}
\For{$r \leftarrow 1$ \textbf{to} $|\mathcal{R}|$}  \label{ln32}
    \State Set $\mathcal{R} \gets \mathcal{R}\setminus\{r\}$ and execute lines \ref{ln1}-\ref{ln31};
    \State Calculate $P_r(x_{ry},\bm{b})$ and $\mathcal{C}_r = C_r + P_r(x_{ry},\bm{b})$;
    \State Set $\mathcal{R} \gets \mathcal{R} \cup \{r\}$;  \Comment{\textit{add $r$ back to $\mathcal{R}$}}
\EndFor         
\State \textbf{return} $x_{ry}$ and $\mathcal{C}_r$;
\end{algorithmic}
\end{algorithm}
\setlength{\textfloatsep}{5pt}

\vspace{-0.5\baselineskip}
\subsection{Heuristic for the VCG Auction-based Resource Allocation}
As \emph{computational efficiency} is an important property for a mechanism design, we design a polynomial-time heuristic to solve the problem $\mathcal{P}_2$ as outlined in Algorithm \ref{alg2} (page 30). We start to iteratively connect each RU $r$ to an Edge/OLT-Cloud $y$. For $r=1$, we initialize the flag $assign \gets 0$, the dummy set $\bar{\mathcal{Y}} \gets \mathcal{Y}$, and find the nearest $y' = \arg\min_y \{D_{ry}\}, y'\in \bar{\mathcal{Y}}$. If the constraints (\ref{eq05}), (\ref{eq09})-(\ref{eq12}) are satisfied for this $y'$, we set $\hat{x}_{ry'}\gets 1$, $t_{y'}\gets 1$, $assign \gets 1$, and calculate the cost of total utilized resources $\mathcal{C}_{tot} = \sum_{y\in\mathcal{Y}} \left(C_\lambda B_y + C_P G_y\right ) t_y$, which is also equal to $\hat{\mathcal{C}}_{r}$, the \emph{shared cost of RU $r$}. If the assignment is not successful, then we remove $y'$ from $\bar{\mathcal{Y}}$ and continue this process until $\bar{\mathcal{Y}} = \emptyset$. In the subsequent iterations, i.e., for $1 < r \leq |\mathcal{R}|$, we find all $y\in \mathcal{Y}$ that satisfy constraints (\ref{eq05}), (\ref{eq09})-(\ref{eq12}) are satisfied for the current RU $r$ and reinitialize the dummy set $\bar{\mathcal{Y}}$. If at least one such $y$ exists, i.e., $|\bar{\mathcal{Y}}| \geq 1$, then we calculate the \emph{dummy costs of total utilized resources} $\mathcal{C}_{tot,y}$ if RU $r$ was connected to each of the Edge/OLT-Cloud $y$. Then we find $y' = \arg\min_y \{\mathcal{C}_{tot,y}\}, y'\in \bar{\mathcal{Y}}$, set $\hat{x}_{ry'}\gets 1$, $t_{y'}\gets 1$, and calculate the updated value of $\mathcal{C}_{tot}$. We also calculate the shared cost of RU $r$ as $[(C_\lambda B_y + C_P G_y)/\sum_r \hat{x}_{ry}]$ for $y=y'$.\par
Once the RU to Edge/OLT-Cloud allocation is complete, we calculate the payment rule. For this purpose, we solve $\mathcal{P}_2$ using steps \ref{ln1}-\ref{ln31} for another $|\mathcal{R}|$ times in the absence of each of the RUs from $\mathcal{R}$. Thus, we can calculate $P_r(\hat{x}_{ry},\bm{b})$, the payment made by each RU $r$ using (\ref{eq18}) and the OPEX of the RU $r$ can be calculated as $\hat{\mathcal{C}}_r = C_r + P_r(\hat{x}_{ry},\bm{b})$. Note that the first for loop iterates for $|\mathcal{R}|$ times to return $\hat{x}_{ry}^*$ and tries to find a suitable Edge/OLT-Cloud $y$ from a set of maximum size $|\mathcal{Y}|$ at every iteration. Therefore, the worst-case time-complexity of this loop is $\mathcal{O}(|\mathcal{R}| \times |\mathcal{Y}|)$. The second for loop iterates for $|\mathcal{R}|$ times to return $\hat{\mathcal{C}}_r$, but it executes the first for loop with complexity $\mathcal{O}(|\mathcal{R}|-1)$. Therefore, the overall time-complexity of Algorithm \ref{alg2} is $\mathcal{O}(|\mathcal{R}|^2 \times |\mathcal{Y}|)$. Also, similar to Algorithm \ref{alg1}, it can shown that the solution produced by Algorithm \ref{alg2} approximates the optimal solution by a factor of $\mathcal{O}(\log_e(|\mathcal{R}|))$.


\vspace{-\baselineskip}
\subsection{Nearest-First (Greedy) and Reinforcement Learning-based Resource Allocation Mechanisms}
To create a performance baseline, we design conventional nearest-first (greedy) and RL-based resource allocation mechanisms. In the \emph{greedy method}, we select the RUs sequentially and attempt to associate them to their nearest Edge/OLT-Cloud subject to the network connectivity (\ref{eq05})-(\ref{eq06}), communication latency (\ref{eq09})-(\ref{eq10}), and processing latency (\ref{eq11})-(\ref{eq12}) constraints, i.e., $y' = \arg\min \{D_{ry} | y'\in\mathcal{Y}, (\ref{eq05})-(\ref{eq06}), (\ref{eq09})-(\ref{eq12})\}$. For each RU $r$, if their nearest Edge/OLT-Cloud $y'$ is unavailable, then we remove $y'$ from the list of Edge/OLT-Clouds and proceed to find the next nearest Edge/OLT-Cloud. In this way, we iteratively try to assign each RU $r$ to an Edge/OLT-Cloud $y$ and $\tilde{x}_{ry}^*$ provides the final allocation. Once the maximum possible RUs are assigned to all Edge/OLT-Clouds, we can calculate the OPEX of the assigned RUs as $\tilde{\mathcal{C}}_r = \sum_{y\in\mathcal{Y}} \left(C_r + C_\lambda \tilde{B}_{ry} + C_P \tilde{G}_{ry} \right) \tilde{x}_{ry}^*$, where the share of throughput and DU-CU function processing resources for each RU $r$ are considered either as proportional sharing:
\begin{align}
    \tilde{B}_{ry} &= \left(\frac{W_r^{UL} B_y^{UL}}{\sum_r x_{ry} W_r^{UL}} \right) + \left(\frac{W_r^{DL} B_y^{DL}}{\sum_r x_{ry} W_r^{DL}} \right), \label{eq20} \\
    \tilde{G}_{ry} &= \left(\frac{\Gamma_r^{UL} G_y^{UL}}{\sum_r x_{ry} \Gamma_r^{UL}}\right)  + \left(\frac{\Gamma_r^{DL} G_y^{DL}}{\sum_r x_{ry} \Gamma_r^{DL}}\right), \label{eq21}
\end{align}
or as uniform sharing:
\begin{align}
    \tilde{B}_{ry} &= \left(\frac{B_y^{UL}}{\sum_r x_{ry}} \right)  + \left(\frac{B_y^{DL}}{\sum_r x_{ry}} \right), \label{eq22} \\
    \tilde{G}_{ry} &= \left(\frac{G_y^{UL}}{\sum_r x_{ry}}\right)  + \left(\frac{G_y^{DL}}{\sum_r x_{ry}}\right). \label{eq23}
\end{align}
\par The OPEX of the unassigned RUs are considered equal to zero. All the steps of this algorithm are summarized in Algorithm \ref{alg3}. We observe that the worst-case time-complexity of the first for loop is $\mathcal{O}(|\mathcal{R}| \times |\mathcal{Y}|)$ and the time-complexity of the second for loop is $\mathcal{O}(|\mathcal{R}|)$. This implies that the overall time-complexity of Algorithm \ref{alg3} is $\mathcal{O}(|\mathcal{R}| \times |\mathcal{Y}|)$.\par
In the \emph{RL-based method}, each RU $r$ randomly selects an Edge/OLT-Cloud $y$ and attempts to establish a front/mid-haul connection by following a \emph{multi-arm bandit algorithm} \cite{Sutton}. The reward of each RU ($\mathfrak{R}_r$) is calculated as the average of $\Delta_r^H$ to front/mid-haul latency ratio and $\Delta_r^{RDC}$ to RU-DU-CU processing latency ratio. The RUs perform both exploration and exploitation ($\epsilon$-greedy with $\epsilon = 0.3$, i.e., explore strategies with probability 0.3, else exploit) to maximize its cumulative reward values and eventually find the best Edge/OLT-Cloud. After RU to Edge/OLT-Cloud allocation is done, the OPEX of the RUs are calculated as above.
\begin{algorithm}[t!]
\caption{Algorithm for nearest-first resource allocation} \label{alg3}
\hspace*{\algorithmicindent} \textbf{Input:} $\mathcal{R}, \mathcal{E}, \mathcal{Y}, D_{ry}, B_y^{U/DL}, W_r^{U/DL}, \Gamma_r^{U/DL}, G_r^{U/DL}$\\
\hspace*{\algorithmicindent} \textbf{Output:} Near-optimal solution: $x_{ry}^*$ and $\mathcal{C}_r^*$\\
\hspace*{\algorithmicindent} \textbf{Initialize:} Sort the elements of $\mathcal{R}$ in the increasing order\\
\hspace*{\algorithmicindent} of ($\max\{W_r^{DL},W_r^{UL}\}$) and ($\max\{\Gamma_r^{DL},\Gamma_r^{UL}\}$) while\\
\hspace*{\algorithmicindent} maintaining an uniform distribution of the percentage of\\ 
\hspace*{\algorithmicindent} ownership of MNOs;
\begin{algorithmic}[1]
\For{$r \leftarrow 1$ \textbf{to} $|\mathcal{R}|$}
    \State Set $\tilde{\mathcal{Y}} \gets \mathcal{Y}$;
    \State Set $assign \gets 0$;
    \While{$assign \neq 1$ \textbf{and} $\tilde{\mathcal{Y}} \neq \emptyset$}
        \State Find $y' = \arg\min_y \{D_{ry}\}, y'\in \tilde{\mathcal{Y}}$;
        \If{constraints (\ref{eq05}), (\ref{eq09})-(\ref{eq12}) are satisfied}
            \State Set $x_{ry'}\gets 1$;
            \State Set $assign \gets 1$;
        \Else
            \State Set $\tilde{\mathcal{Y}} \gets \tilde{\mathcal{Y}}\setminus\{y'\}$;
        \EndIf
    \EndWhile
\EndFor
\For{$r \leftarrow 1$ \textbf{to} $|\mathcal{R}|$}
    \If{$\sum_y x_{ry} = 1$}
        \State Calculate $\tilde{B}_{ry}$, $\tilde{G}_{ry}$, and $\mathcal{C}_r$;
    \Else
        \State Set $\tilde{B}_{ry} = 0$, $\tilde{G}_{ry} = 0$, and $\mathcal{C}_r = 0$;
    \EndIf
\EndFor
\State \textbf{return} $x_{ry}$ and $\mathcal{C}_r$;
\end{algorithmic}
\end{algorithm}
\setlength{\textfloatsep}{5pt}
\section{Results and Discussions} \label{sec6}
To evaluate our proposed frameworks, we consider a multi-tenant O-RAN deployment area of dimension $5\times 5$ km$^2$. In this area, 8 macro-cell RUs (coverage = 1 km) and 30 small-cell RUs (coverage = 0.5 km) from three different MNOs coexist (refer to Appendix \ref{sec4a}). The front/mid-haul datarate and RU-DU-CU processing efforts of the RUs vary over time depending on the RU configurations and the chosen split option. For example, if the RUs are configured with 2x2 MIMO, 2 layers, 50 MHz bandwidth, 15 kHz sub-carrier spacing, MCS index 16, and slot duration = 0.5 msec, the maximum uplink datarate with Split-7.2 is 2.304 Gbps, and the maximum downlink datarate with Split-7.3 is 0.432 Gbps. Each radio head's total RU-DU-CU processing requirement is nearly 550 GOPS/slot. We consider that in each radio head, 25\% resources are used for uRLLC services and 75\% resources are used for eMBB services. The one-way front-haul link latency bound is $\sim$100 $\mu$sec and we choose the RU-DU-CU processing latency bound 325 $\mu$sec for uRLLC services and 975 $\mu$sec for eMBB services \cite{delay_calc}. The reduced queueing latency of the uplink data at the ONUs is 15 $\mu$sec as Co-DBA is used for uplink transmission \cite{5g_fh_bw} and the data are transmitted as periodic bursts of duration 31.25 $\mu$sec \cite{pon_oran}. Each wavelength of the TWDM-PON can support a throughput of 25 Gbps and we can aggregate multiple such wavelengths to achieve a few hundred Gbps of total throughput. A group of RUs are connected to a level-1 reflective splitter (locations found through $k$-means clustering) and a few level-1 reflective splitters are connected to a level-2 reflective splitter located at the center of the area. We assume that the RUs can be connected to the OLT-Clouds via the North-South links and to the Edge-Clouds via creating East-West virtual-PON links. We arbitrarily assume that each RU pays a default cost of \euro 100/day to the NSP (other network economic pricing schemes can also be employed but beyond the scope of this paper), although it has very little effect on the results as it is not associated with any decision variables. In addition, the cost for throughput used is \euro 0.5/Gbps \cite{fib_rent}, and the cost for leasing DU/CU resources is \euro 1.5/GOPS \cite{F_split}.\par
\vspace{-\baselineskip}
\begin{figure}[!t]
  \centering
  \subfloat[Min-max fairness vs. baseline algorithms]{%
    \includegraphics[width=0.9\columnwidth]{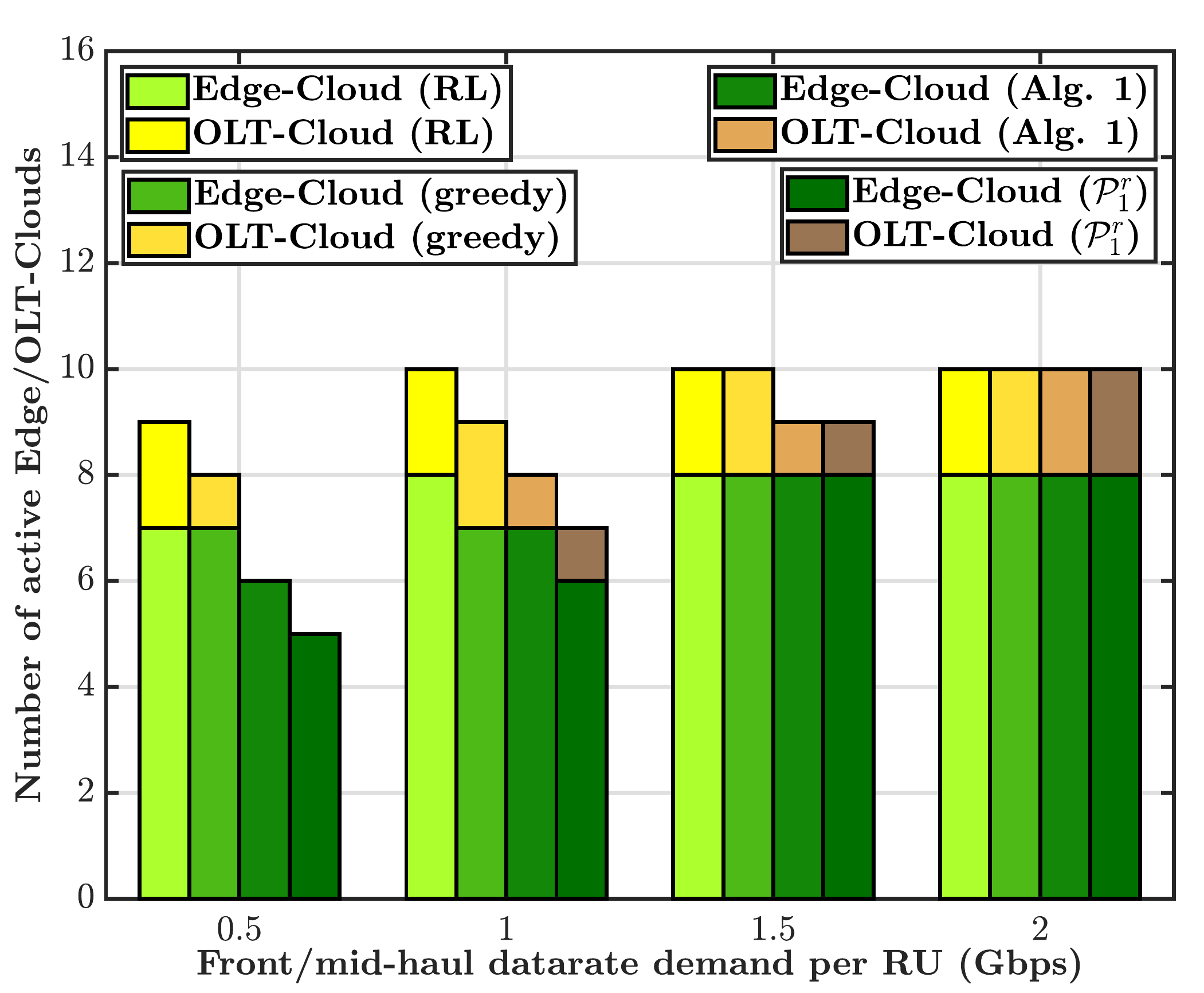}\label{ec1}%
  }
  
  \subfloat[VCG auction vs. baseline algorithms]{%
    \includegraphics[width=0.9\columnwidth]{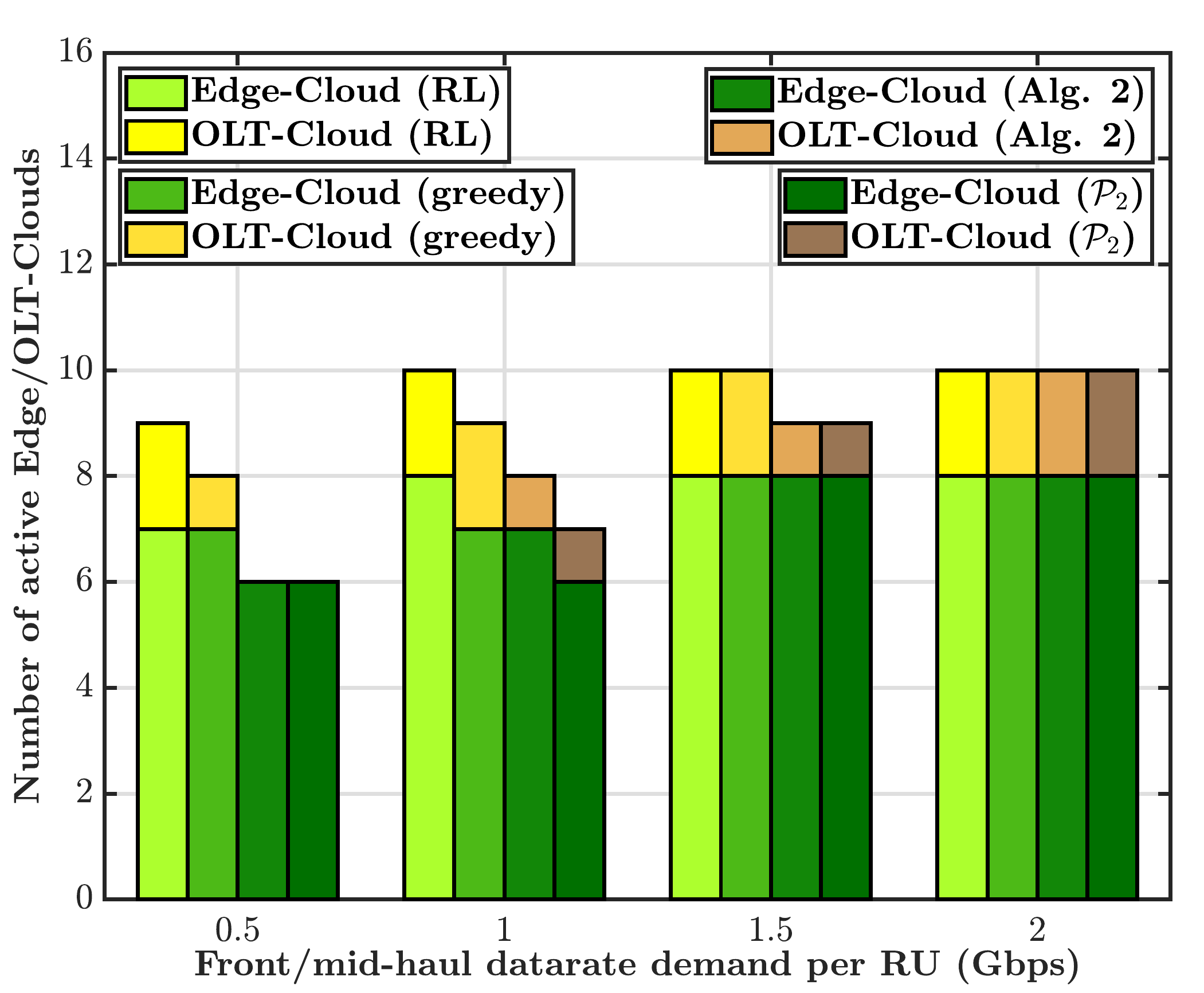}\label{ec2}%
  }

  \caption{Comparison of the total number of active Edge/OLT-Clouds obtained by the optimal solution, heuristics, and baseline methods with more computational resources at Edge-Clouds than at OLT-Clouds.}
  \label{cloud_compare}
\end{figure}
\setlength{\textfloatsep}{5pt}
\begin{table}[!b]
\centering
\caption{Network Evaluation Scenarios}
\label{table2}
\resizebox{\columnwidth}{!}{%
\begin{tabular}{lcc|cc|cc}
\cline{2-7}
\multirow{2}{*}{}                                                                           & \multicolumn{2}{c|}{\textbf{Scenario - I}} & \multicolumn{2}{c|}{\textbf{Scenario - II}} & \multicolumn{2}{c}{\textbf{Scenario - III}} \\ \cline{2-7} 
                                                                                            & \multicolumn{1}{c|}{\textbf{Edge}}  & \textbf{OLT}    & \multicolumn{1}{c|}{\textbf{Edge}}    & \textbf{OLT}  & \multicolumn{1}{c|}{\textbf{Edge}} & \textbf{OLT}    \\ \hline
\multicolumn{1}{c|}{\textbf{\begin{tabular}[c]{@{}c@{}}Maximum \\ GOPS/slot\end{tabular}}}   & \multicolumn{1}{c|}{$1.5\times10^4$}  & $4.5\times10^4$ & \multicolumn{1}{c|}{$3\times10^4$}  & $3\times10^4$ & \multicolumn{1}{c|}{$4.5\times10^4$} & $1.5\times10^4$ \\ \hline
\multicolumn{1}{c|}{\textbf{\begin{tabular}[c]{@{}c@{}}Maximum \\ throughput\end{tabular}}} & \multicolumn{1}{c|}{50 Gbps}        & 600 Gbps        & \multicolumn{1}{c|}{100 Gbps}         & 400 Gbps      & \multicolumn{1}{c|}{150 Gbps}      & 200 Gbps        \\ \hline
\end{tabular}
}
\end{table}
\setlength{\textfloatsep}{10pt}
\begin{figure*}[!t]
  \centering
  \subfloat[Scenario-I]{%
    \includegraphics[width=0.333\textwidth,height=5cm]{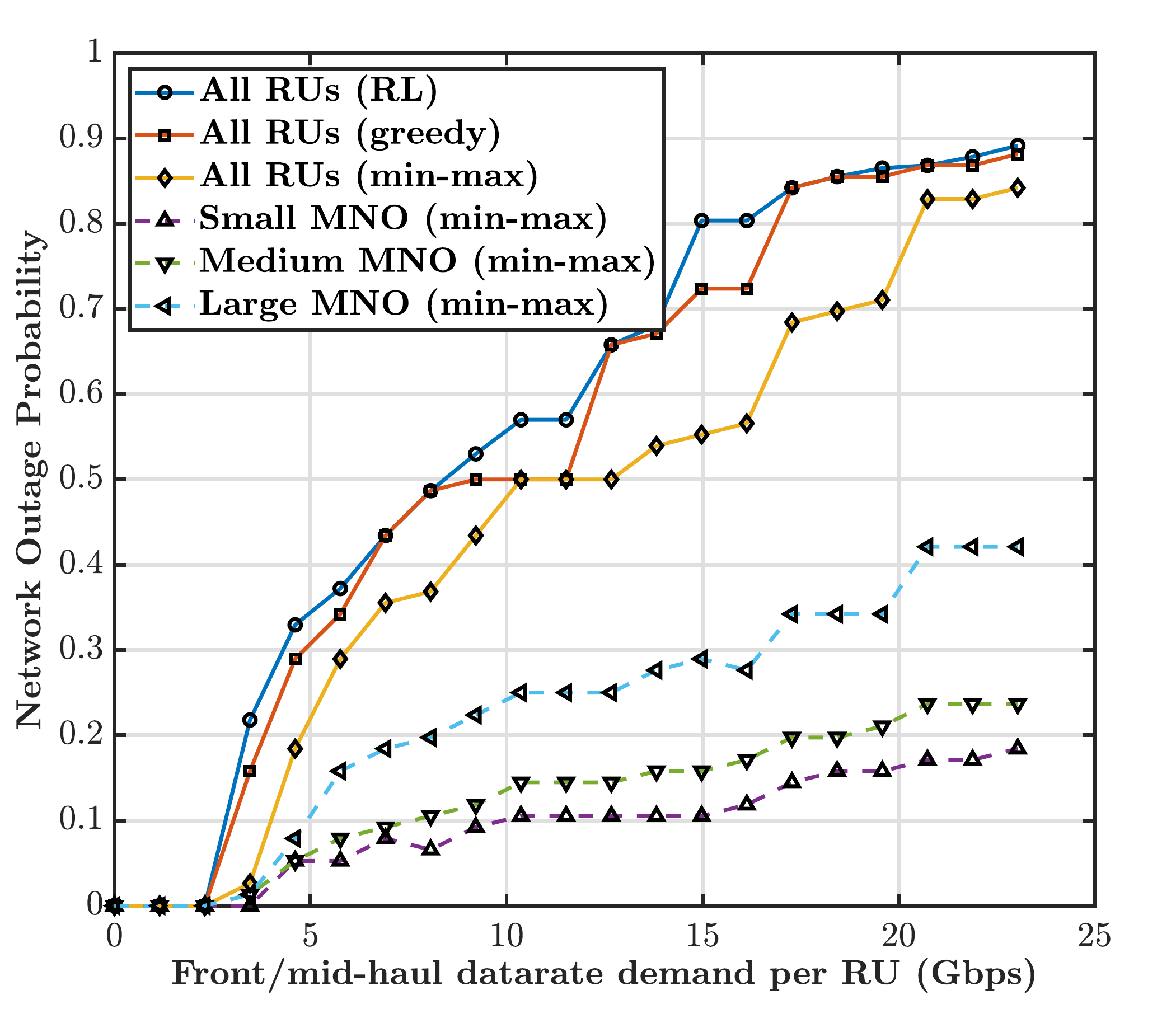}\label{outage1}%
  }
  \subfloat[Scenario-II]{%
    \includegraphics[width=0.333\textwidth,height=5cm]{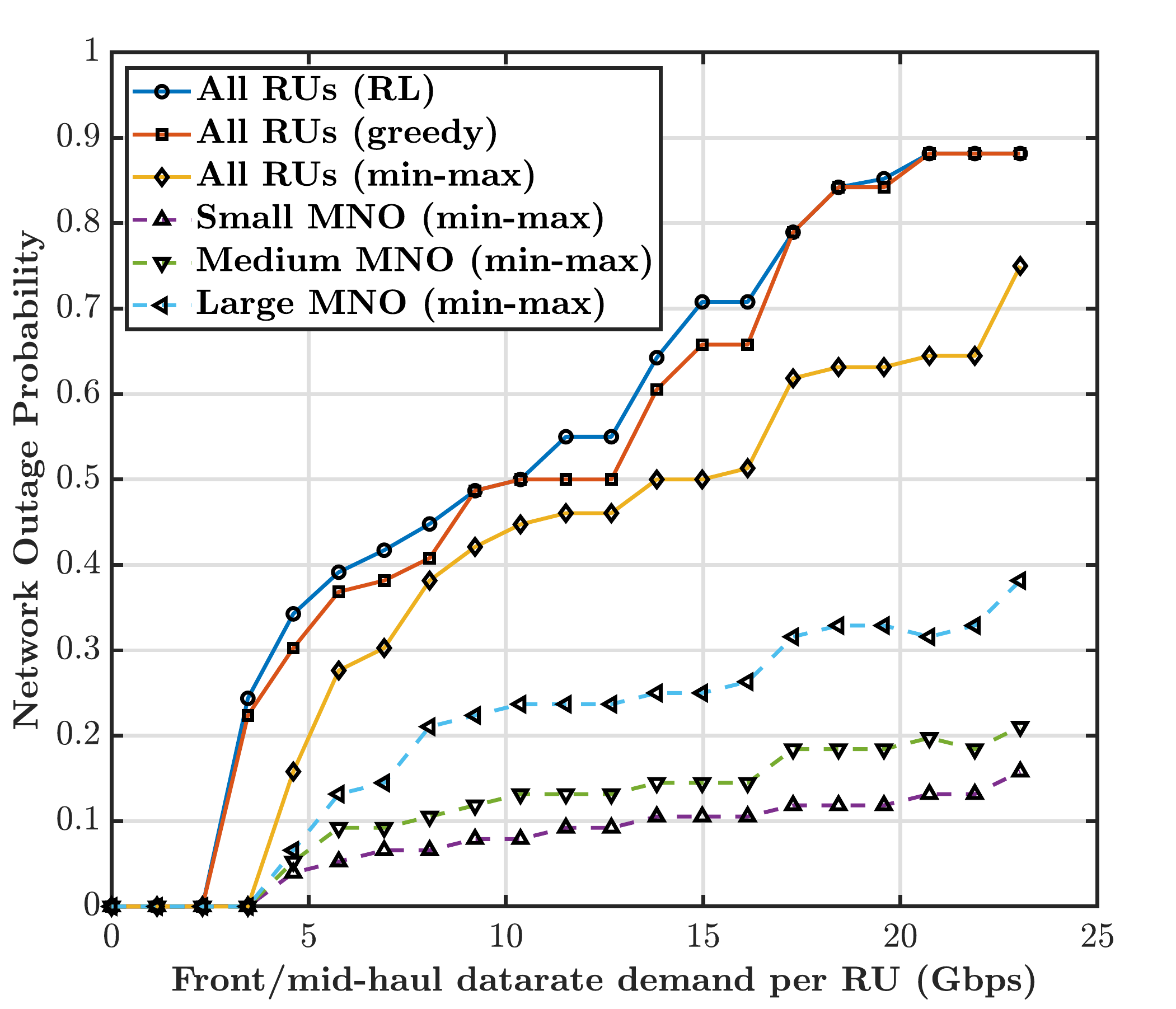}\label{outage2}%
  }
  \subfloat[Scenario-III]{%
    \includegraphics[width=0.333\textwidth,height=5cm]{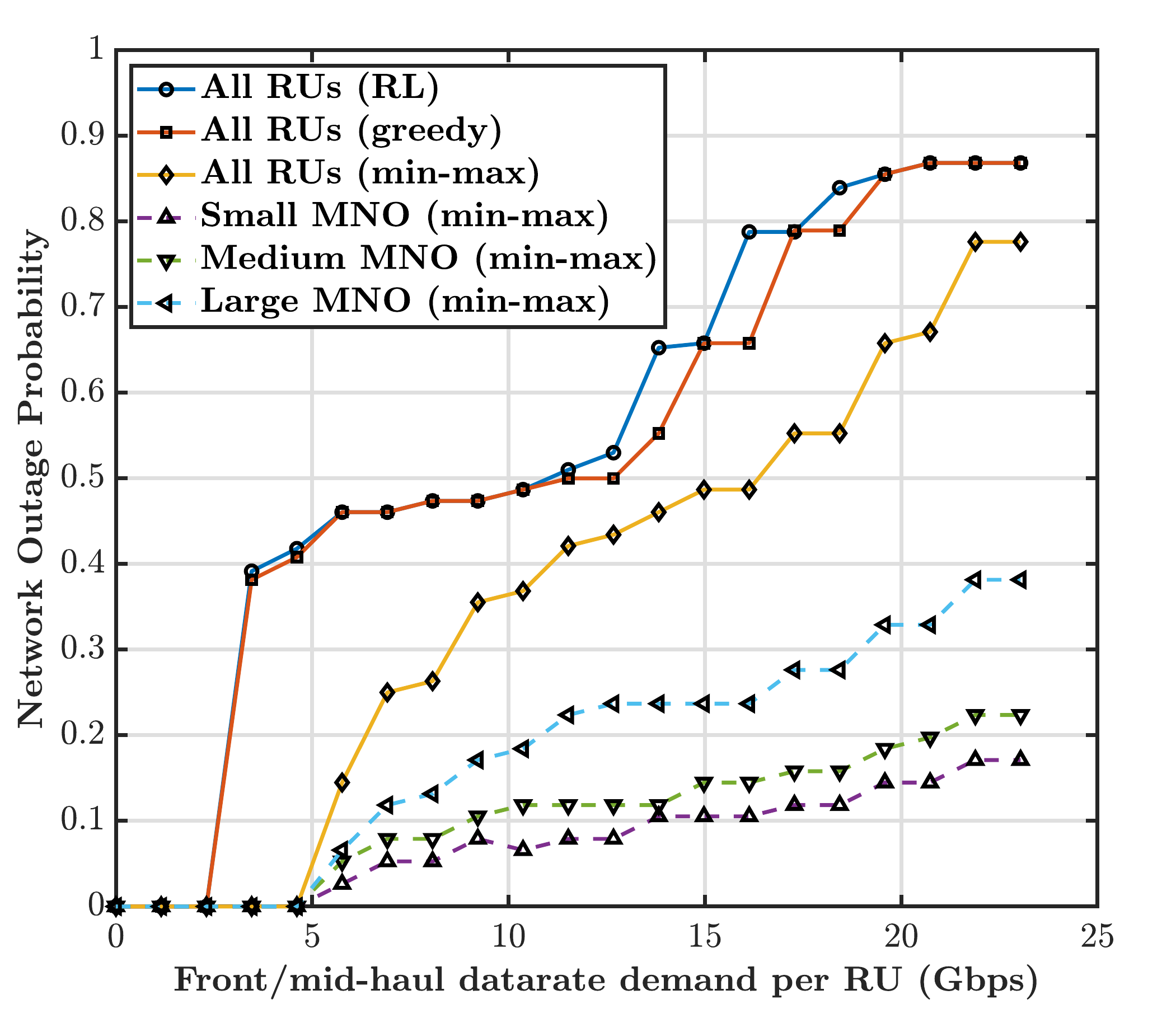}\label{outage3}%
  }

  \caption{Comparison of overall O-RAN outage probability and outage probability for small, medium, and large MNOs against increasing front/mid-haul datarate demand per RU with different Edge/OLT-Cloud resource distributions obtained by min-max fairness and baseline (greedy nearest-first and reinforcement learning-based) algorithms.}
  \label{outage_compare1}
\end{figure*}
\setlength{\textfloatsep}{1pt}

\begin{figure*}[!t]
  \centering
  \subfloat[Scenario-I]{%
    \includegraphics[width=0.333\textwidth,height=5cm]{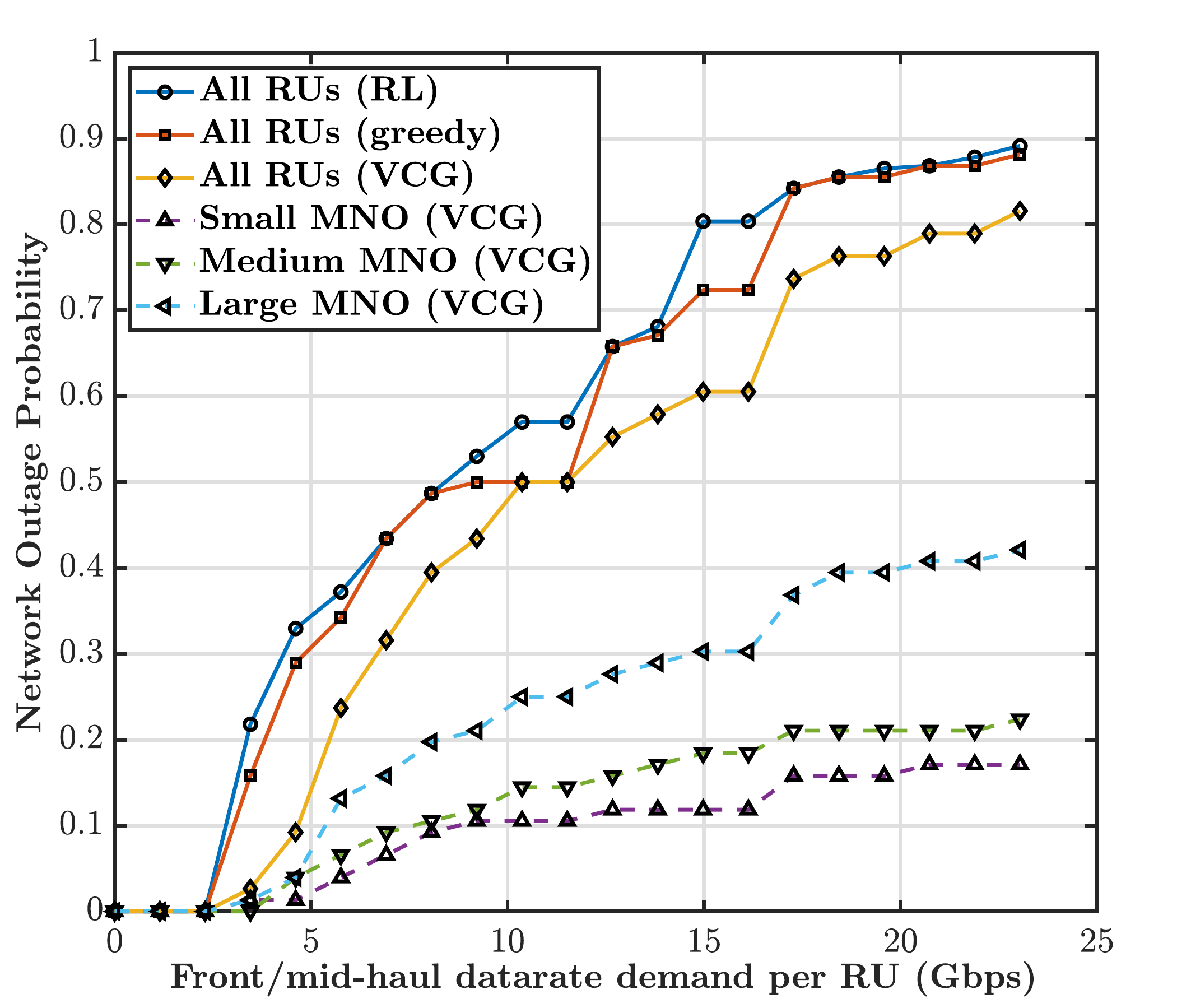}\label{outage4}%
  }
  \subfloat[Scenario-II]{%
    \includegraphics[width=0.333\textwidth,height=5cm]{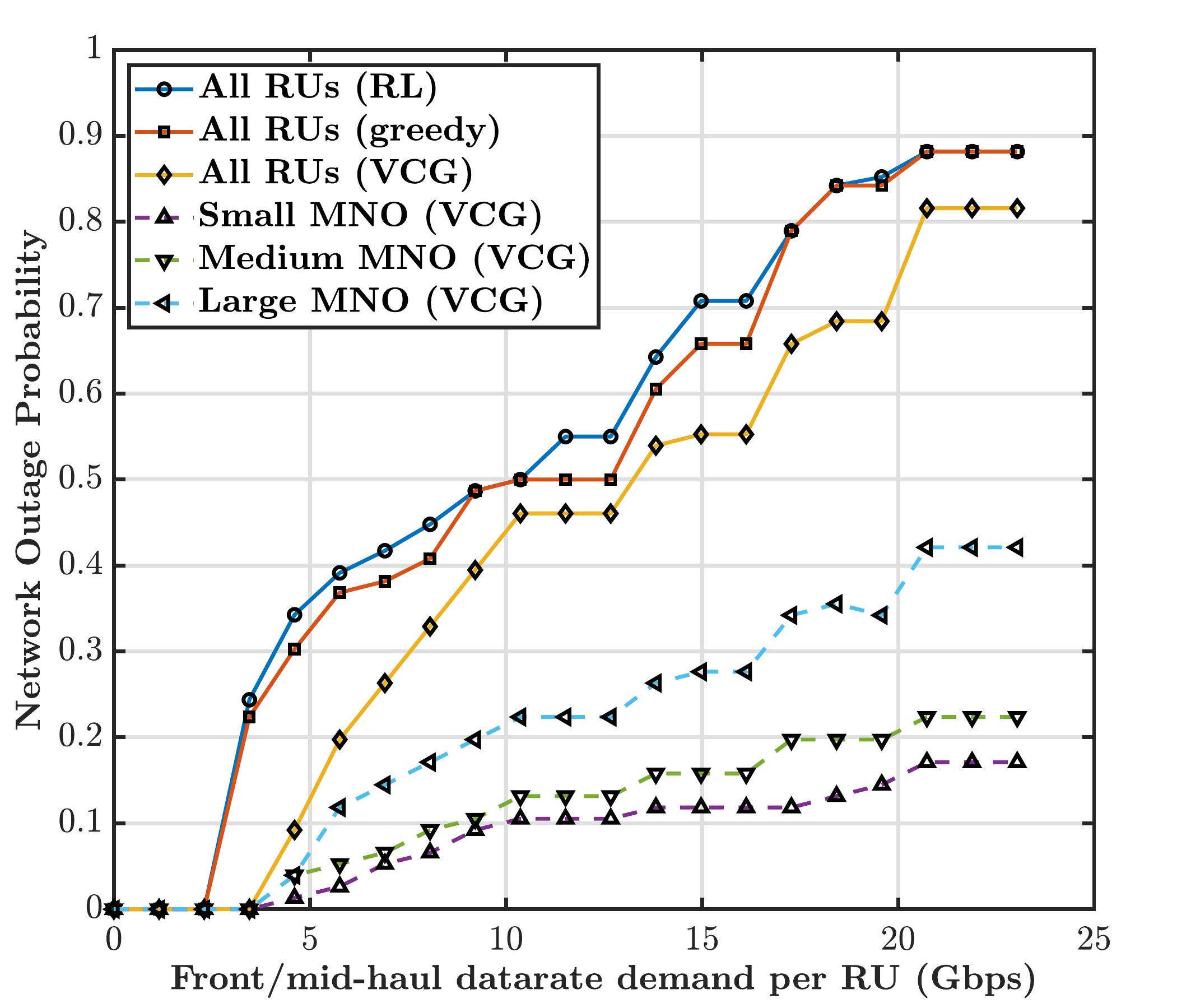}\label{outage5}%
  }
  \subfloat[Scenario-III]{%
    \includegraphics[width=0.333\textwidth,height=5cm]{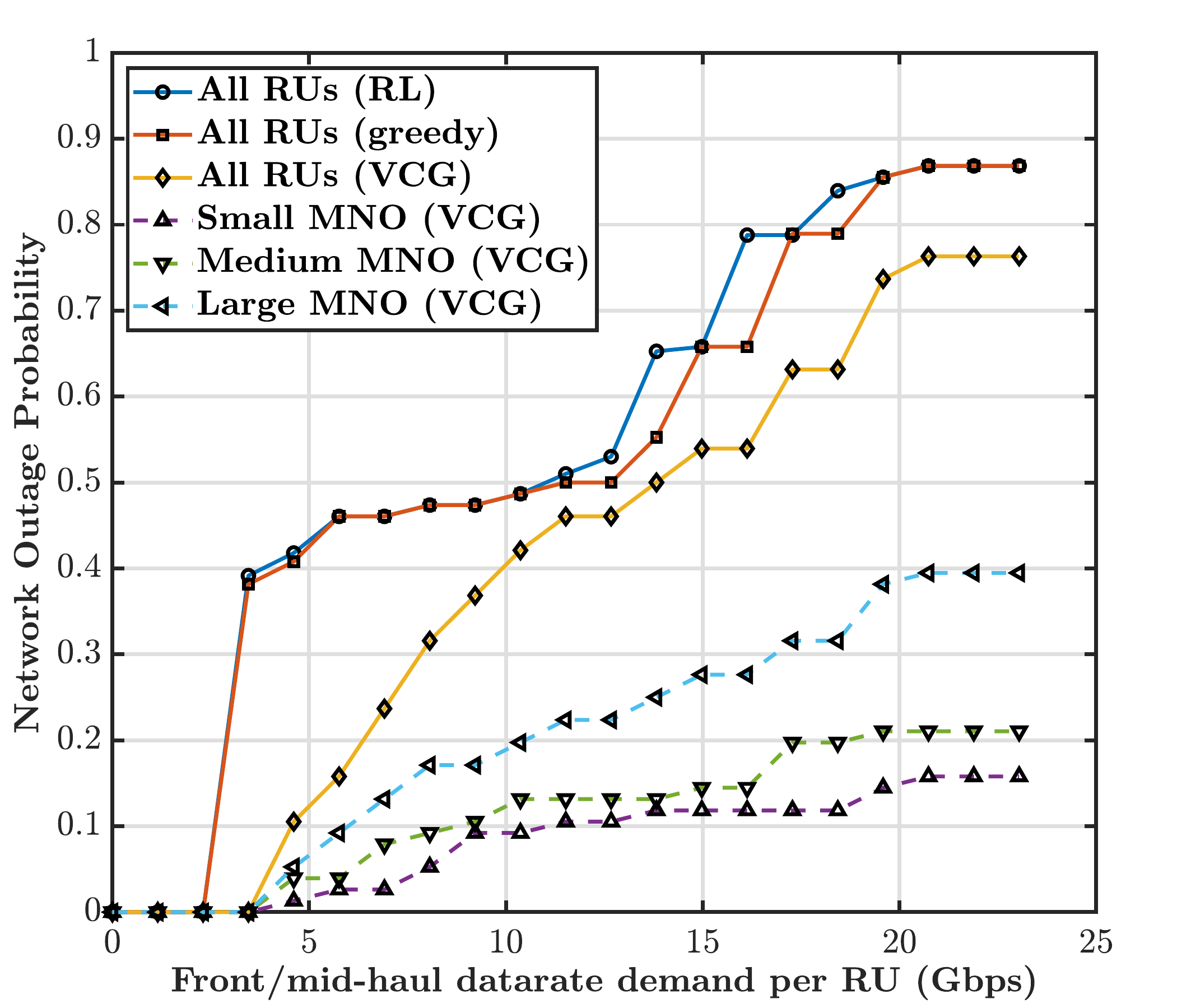}\label{outage6}%
  }

  \caption{Comparison of overall O-RAN outage probability and outage probability for small, medium, and large MNOs against increasing front/mid-haul data per RU with different Edge/OLT-Cloud resource distributions obtained by VCG auction-based and baseline (greedy nearest-first and reinforcement learning-based) algorithms.}
  \label{outage_compare2}
\end{figure*}
\setlength{\textfloatsep}{1pt}
Firstly, we compare the optimal solutions of the min-max fairness and VCG auction-based resource allocation methods with their corresponding heuristics. In this evaluation, each Edge-Cloud has a maximum capacity of $4.5\times10^4$ GOPS/slot and each OLT-Cloud has a maximum capacity of $1.5\times10^4$ GOPS/slot. These values are chosen such that a feasible benchmark solution can be obtained by the greedy baseline algorithm with front/mid-haul datarate demand upto 2 Gbps per RU. We use OCTERACT, a global optimal mixed-integer nonlinear programming solver integrated with the AMPL platform to evaluate the optimal solutions of our formulated INLPs in a computer with Intel Core i7 processor and 32 GB RAM. In Fig. \ref{ec1}, we compare the optimal solution of $\mathcal{P}_1^r$ against the solutions obtained using Algorithm \ref{alg1}, the greedy baseline Algorithm \ref{alg3}, and the RL-based method. Again, in Fig. \ref{ec2}, we compare the optimal solution of $\mathcal{P}_2$ against the solutions obtained using Algorithm \ref{alg2}, the greedy baseline Algorithm \ref{alg3}, and the RL-based method with the same dataset. From these figures, we observe that the optimal solution of $\mathcal{P}_1^r$ is the highest performing during lower load conditions. The nearest-first method makes several sub-optimal RU to Edge/OLT-Cloud assignments as it connects each RU to its nearest (based on intermediate North-South or East-West link distance) Edge/OLT-Cloud if the latency constraints are satisfied. The RL-based method requires less computation than other methods, but each RU independently attempts to connect to some Edge/OLT-Cloud that gives a higher reward, in turn lower latency values. Thus, the RU to Edge/OLT-Cloud assignment does not incorporate minimization of overall network resource utilization. However, all the solutions at medium and higher load conditions become similar as network load increases, because a large percentage of the potential Edge/OLT-Clouds gets activated and the scope of sub-optimal RU to Edge/OLT-Cloud assignments is reduced.\par
\begin{figure*}[!t]
  \centering
  \subfloat[Scenario-I]{%
    \includegraphics[width=0.333\textwidth,height=5cm]{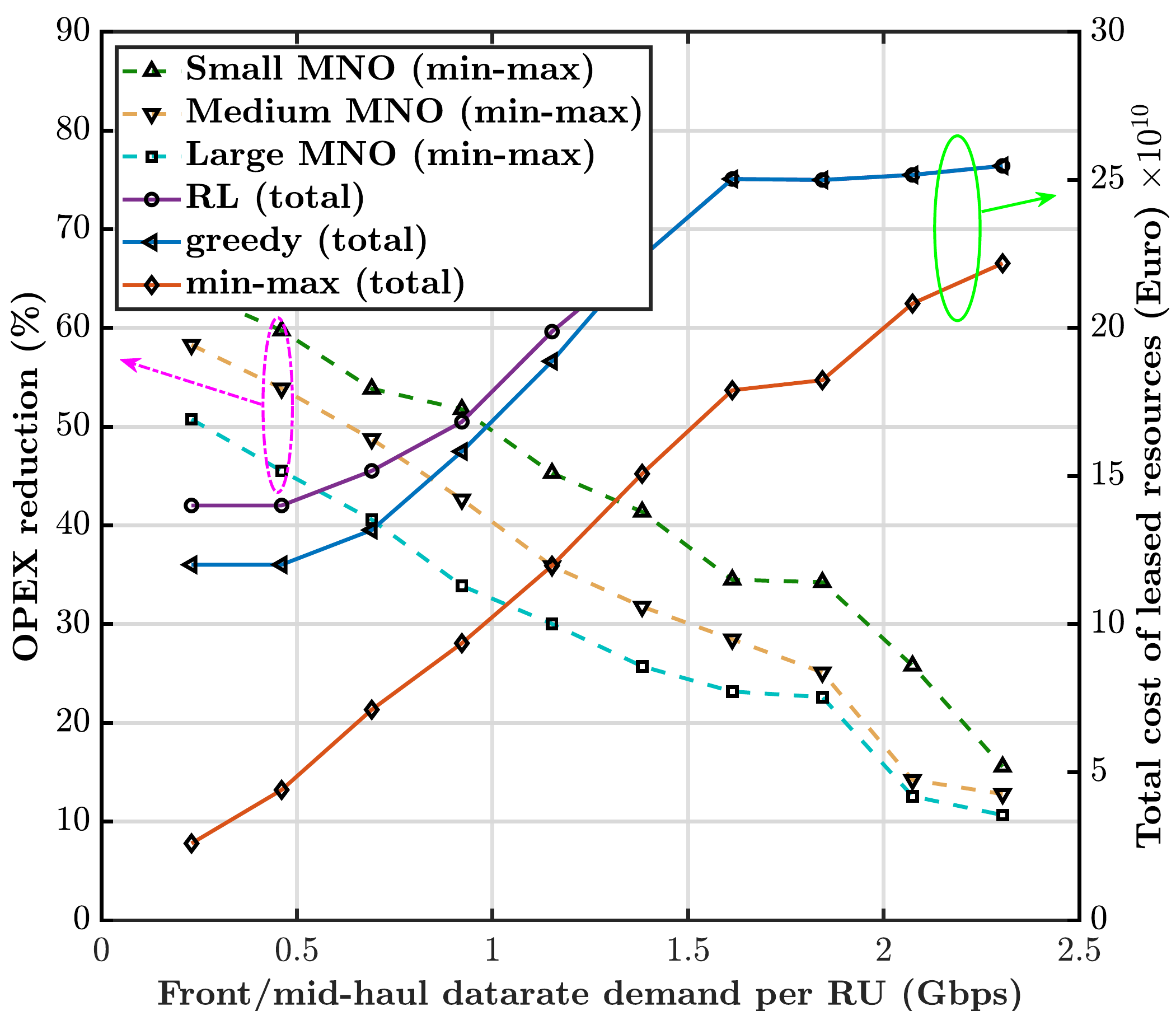}\label{cost1}%
  }
  \subfloat[Scenario-II]{%
    \includegraphics[width=0.333\textwidth,height=5cm]{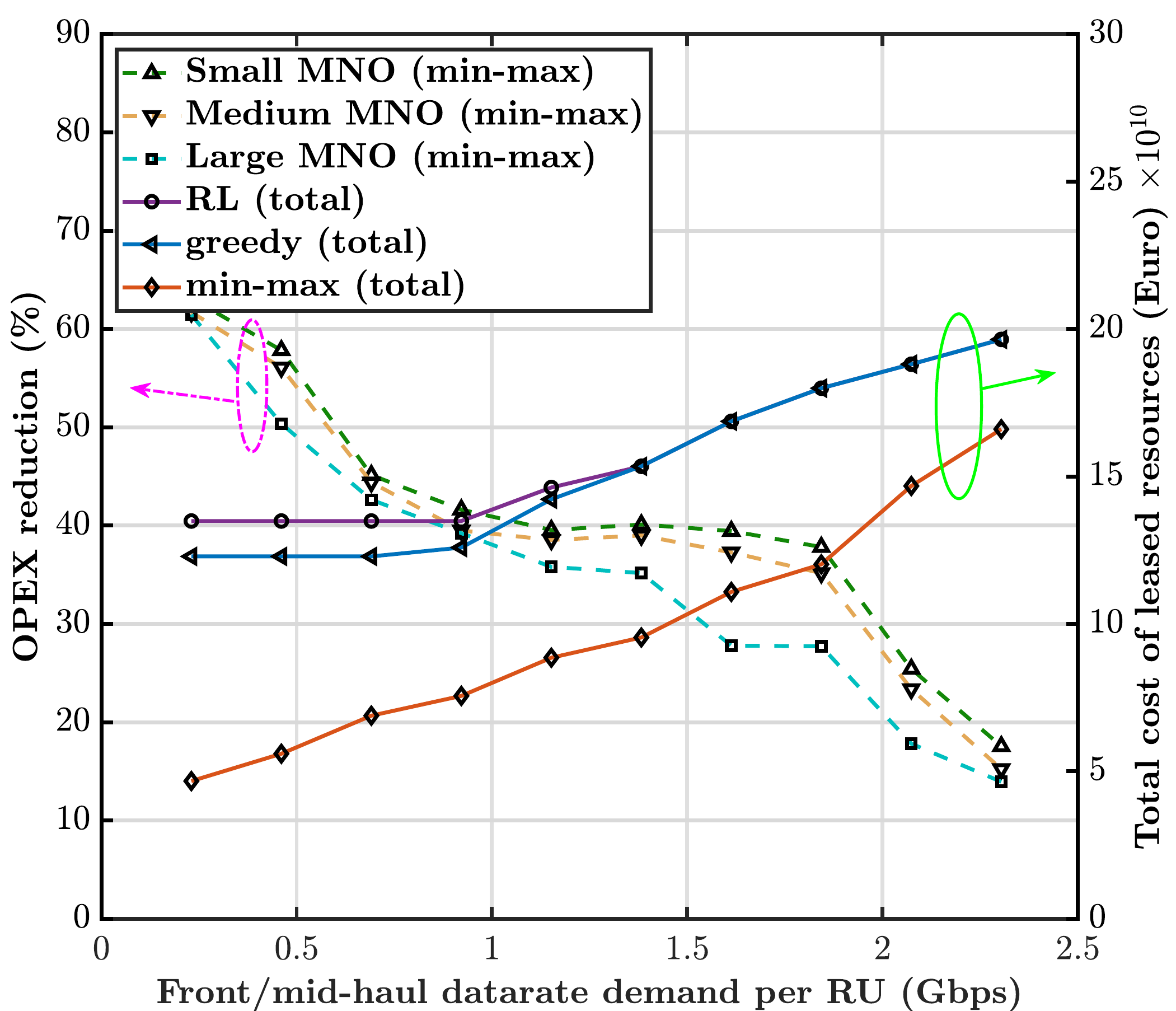}\label{cost2}%
  }
  \subfloat[Scenario-III]{%
    \includegraphics[width=0.333\textwidth,height=5cm]{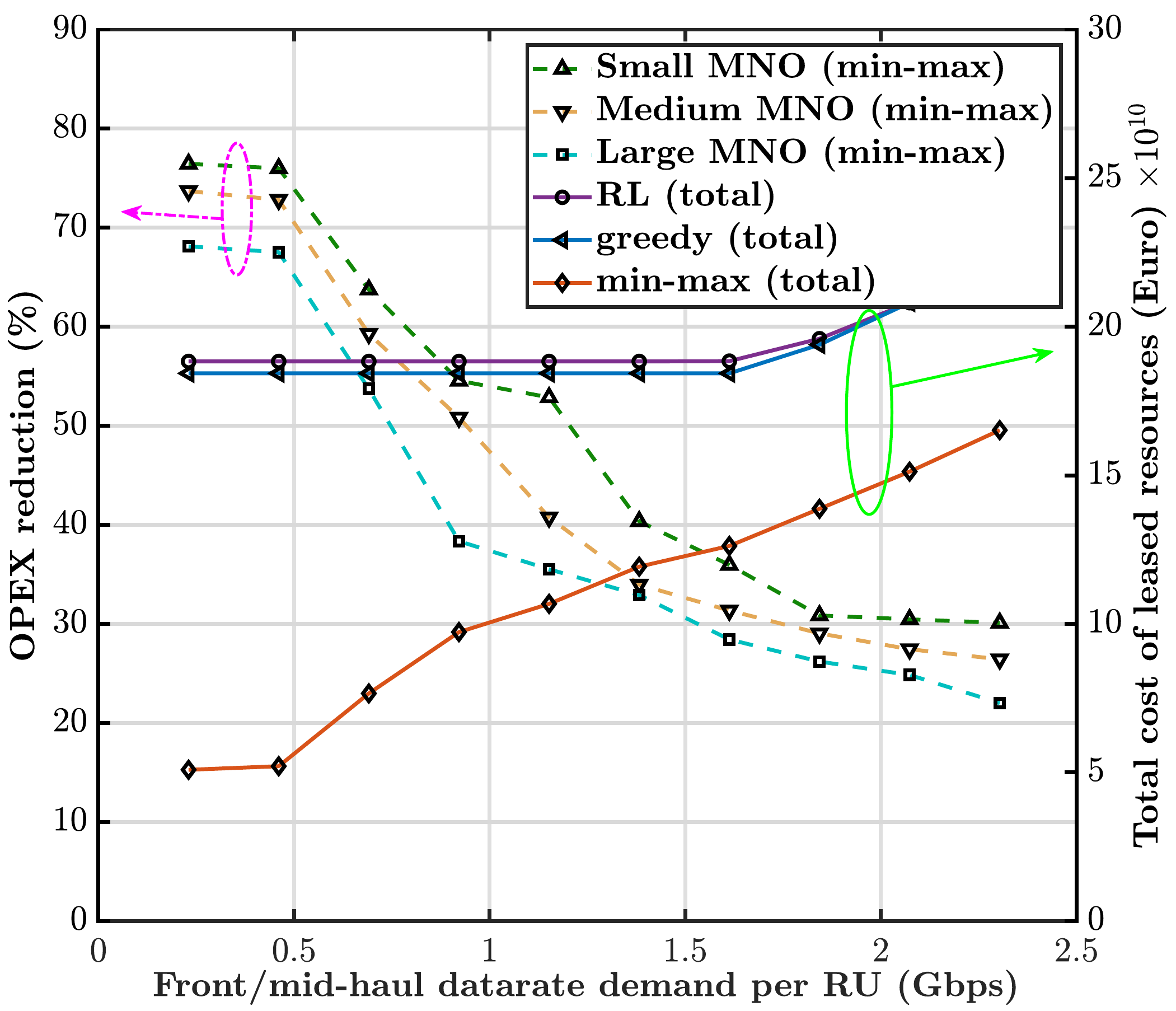}\label{cost3}%
  } 

  \caption{Comparison of the total cost of leased resources (\euro) and OPEX reduction percentage for small, medium, and large MNOs against network load variation with different Edge/OLT-Cloud resource distributions obtained by min-max fairness and baseline (greedy nearest-first and reinforcement learning-based) algorithms.}
  \label{cost_compare1}
\end{figure*}
\setlength{\textfloatsep}{1pt}

\begin{figure*}[!t]
  \centering
  \subfloat[Scenario-I]{%
    \includegraphics[width=0.333\textwidth,height=5cm]{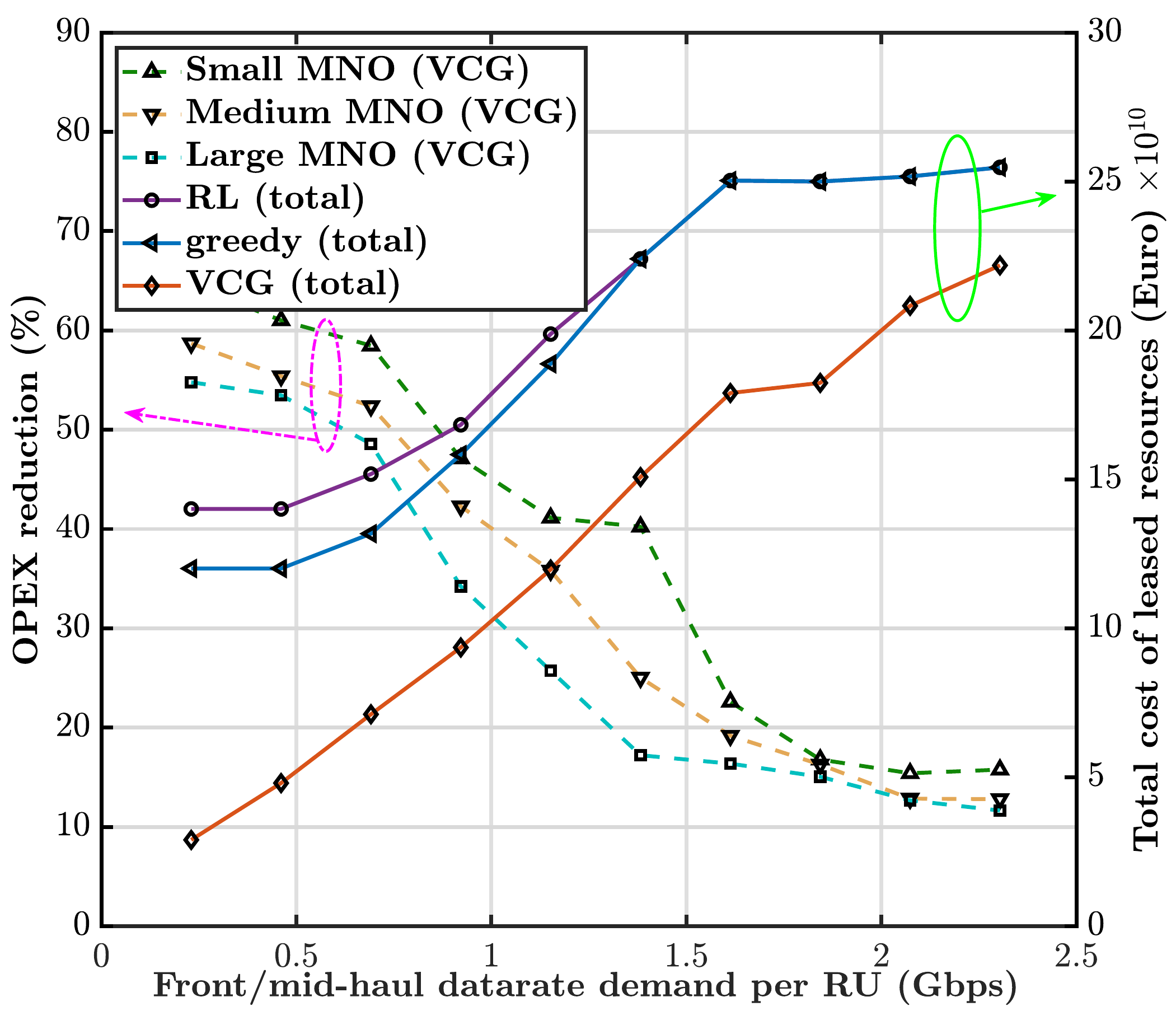}\label{cost4}%
  }
  \subfloat[Scenario-II]{%
    \includegraphics[width=0.333\textwidth,height=5cm]{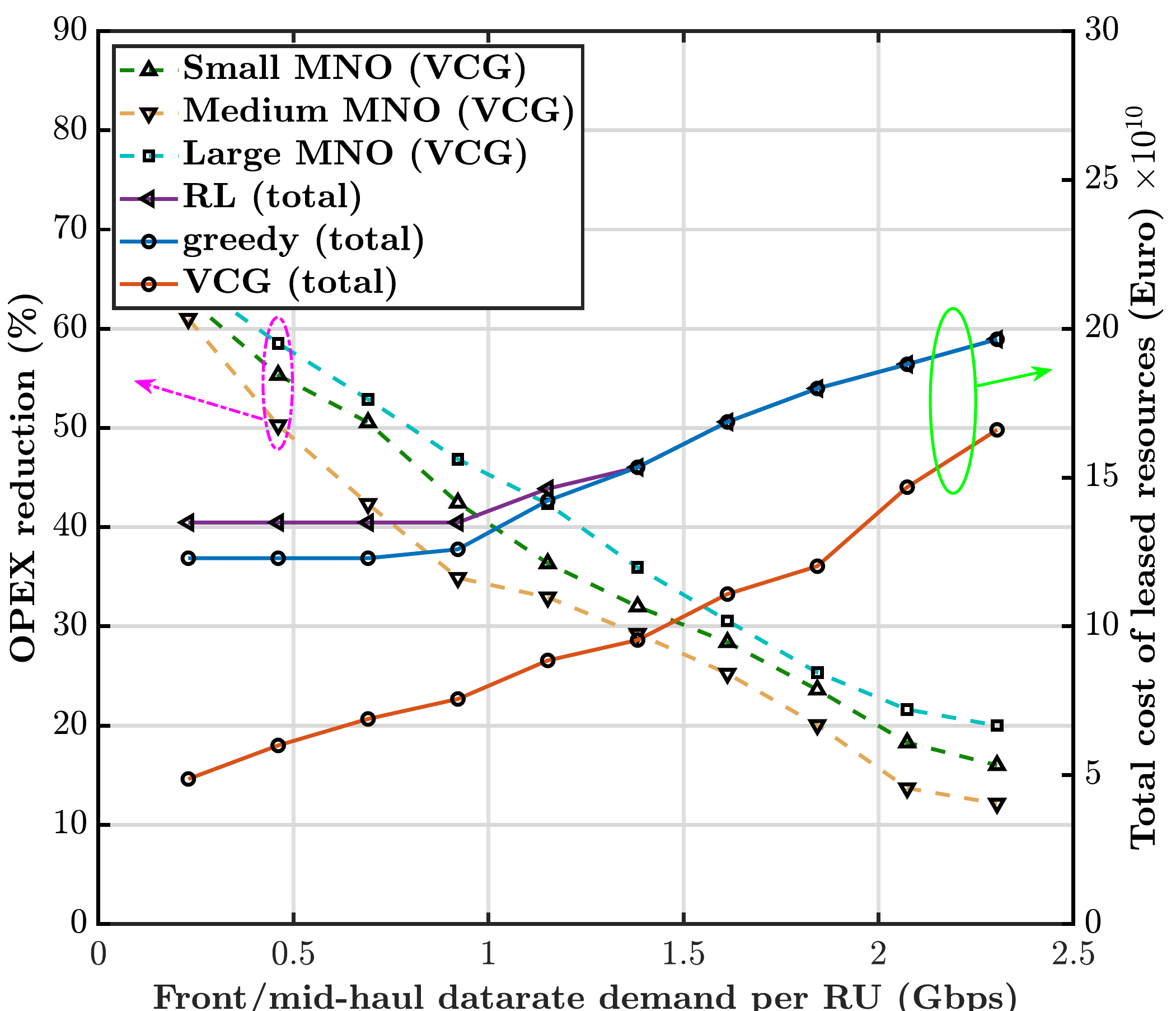}\label{cost5}%
  }
  \subfloat[Scenario-III]{%
    \includegraphics[width=0.333\textwidth,height=5cm]{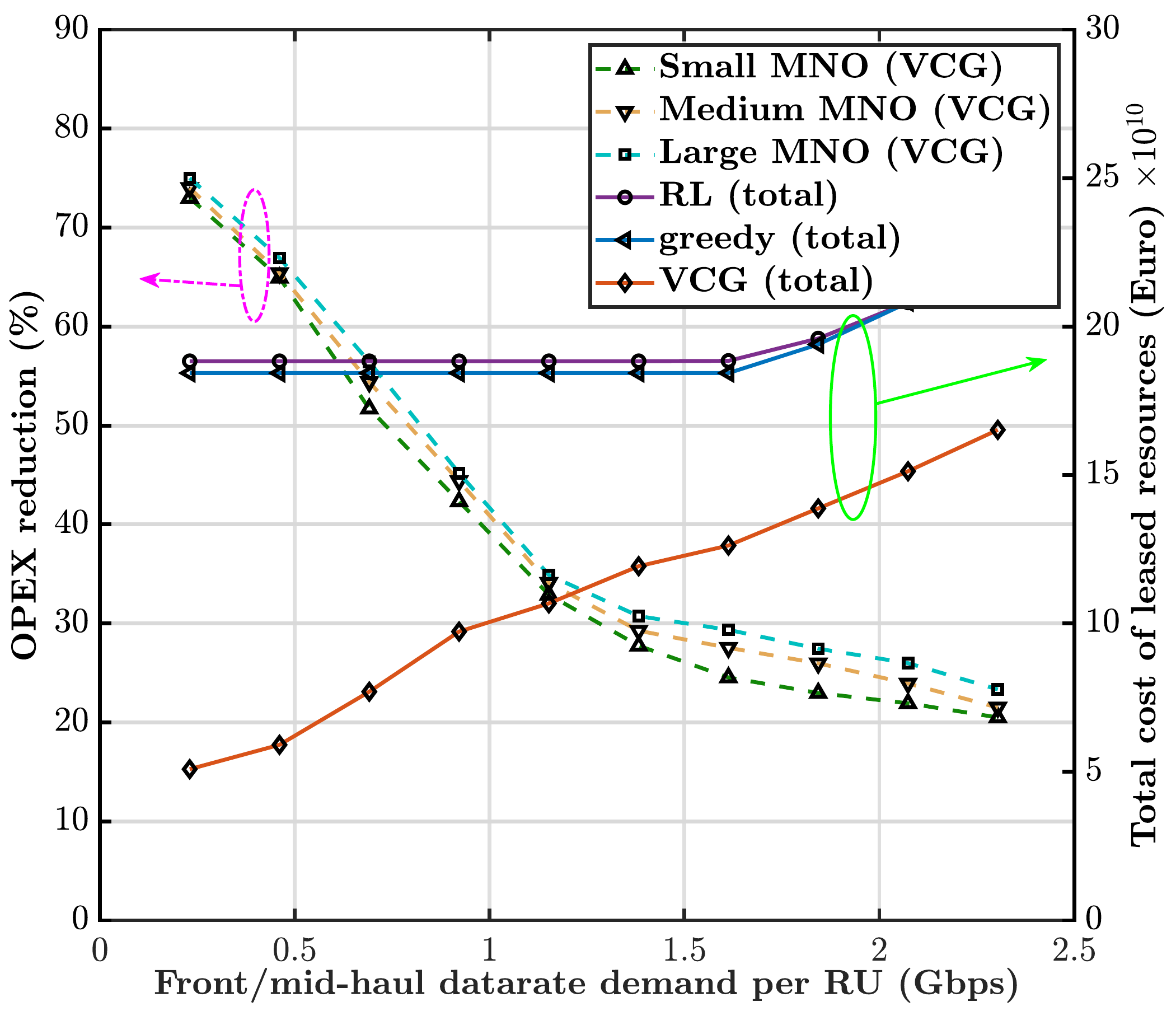}\label{cost6}%
  }

  \caption{Comparison of the total cost of leased resources (\euro) and OPEX reduction percentage for small, medium, and large MNOs against network load variation with different Edge/OLT-Cloud resource distributions obtained by VCG auction-based and baseline (greedy nearest-first and reinforcement learning-based) algorithms.}
  \label{cost_compare2}
\end{figure*}
\setlength{\textfloatsep}{1pt}
Next, we compare the \emph{network outage probabilities} (defined as the ratio of the total number of RUs that could not be connected to some Edge/OLT-Cloud to the total number of RUs present in the network) achieved by our proposed methods. We consider that the RUs are owned by three different MNOs where a small MNO-1 has 20\%, a medium MNO-2 has 30\%, and a large MNO-3 has 50\% ownership. Additionally, we consider three scenarios with different computational resource distributions among Edge-Clouds and OLT-Clouds as outlined in Table \ref{table2}. Figs. \ref{outage1}-\ref{outage3} show the overall O-RAN outage probabilities obtained through the min-max fairness and baseline (greedy nearest-first and RL-based) algorithms. Similarly, Figs. \ref{outage4}-\ref{outage6} show the overall O-RAN outage probabilities obtained through the VCG auction-based and baseline (greedy nearest-first and RL-based) algorithms. We observe that all the methods show zero outage probability as long as there are sufficient front/mid-haul and Edge/OLT-Cloud resources to accommodate all the RUs. However, network outage probability increases fastest with the RL-based method, followed by the nearest-first baseline method due to their inefficient utilization of network resources, but the outage probability increases relatively slower with the VCG auction-based method and slowest with the min-max fairness method. We also plot the outage probabilities of the small, medium, and large MNOs, which show that the outage probabilities of the small, medium, and large MNOs vary according to their percentage of ownership, i.e., the highest for the large MNO and smallest for the small MNO, in all scenarios.\par
%
Finally, we compare the total cost of leased front/mid-haul and Edge/OLT-Cloud resources obtained by the min-max fairness and baseline (greedy nearest-first and RL-based) algorithms. In Figs. \ref{cost1}-\ref{cost3}, we observe that during the low-load conditions, the performance of the min-max fairness approach is significantly better than the baseline (greedy nearest-first and RL-based) methods. However, as the front/mid-haul datarate demand per RU increases, the performance of all algorithms becomes similar due to the saturation of available resources. Note that the total cost of leased resources (right y-axis) increases as we place more resources at the Edge-Clouds. Also, the percentage of OPEX savings decreases (left y-axis) as resource utilization increases for all the MNOs. In all scenarios, \emph{the OPEX saving percentage is highest for the small MNO and lowest for the large MNOs.} Similarly, Figs. \ref{cost4}-\ref{cost6} compare the total cost of leased resources (right y-axis) obtained and OPEX reduction percentages (left y-axis) by the VCG auction-based and baseline (greedy nearest-first and RL-based) algorithms. We observe that the total cost of leased resources obtained by the VCG auction-based algorithm is very similar to the min-max fairness algorithm, but the OPEX saving percentages for small and medium MNOs are not always lower than the large MNO because the cost of the leased resources is uniformly shared among the RUs. Thus, often the small and medium MNOs pay a much higher price than their actual resource requirements. These results justify that our proposed min-max fair resource allocation framework creates a multi-tenant O-RAN ecosystem that is sustainable for small, medium, and large MNOs.\par

\section{Conclusion} \label{sec7}
In this paper, we have proposed a multi-tenant O-RAN architecture where RUs from multiple MNOs can lease resources for their DU-CU function processing from Edge/OLT-Clouds over TWDM-PON-based open front/mid-haul interfaces. These TWDM-PONs use reflective splitters to facilitate East-West communication links along with traditional North-South communication links for better mesh connectivity among the RUs and Edge/OLT-Clouds. We have proposed two methods for allocating front/mid-haul and DU-CU processing resources to RUs based on min-max fairness and VCG auction mechanisms. In turn, we have formulated the corresponding INLPs and have designed time-efficient heuristic algorithms. Through numerical evaluation, we have investigated the performance of these frameworks against different distributions of cloud resources at Edge and OLT locations. \textcolor{black}{We have shown that both min-max fairness and VCG auction-based methods achieve a much lower network outage probability than the baseline (greedy nearest-first and RL-based) methods due to the efficient utilization of network resources}. Note that the VCG auction-based method can ensure the extraction of truthful information from the RUs, but the min-max fairness method ensures that the OPEX of the RUs are proportional to their actual resource requirements. Moreover, we have shown that the min-max fairness method can reduce the OPEX of all MNOs by more than 20\% (at high load) to nearly 75\% (at low load). We believe that our proposed frameworks have successfully laid cornerstones for developing further O-RAN resource allocation strategies.

\appendices
\section{Proof of Proposition 1} \label{sec1a}
\begin{IEEEproof}
We observe that if our main problem formulation $\mathcal{P}_1$ has an optimal solution $x_{ry}^*$, then our reformulated problem $\mathcal{P}_1^r$ also has an optimal solution $(x_{ry}^*,M^*)$. This implies that we have found the best possible $M^*$ as the maximum value of $\sum_{y\in\mathcal{Y}} \left(C_r + C_\lambda B_{ry} + C_P G_{ry}\right ) x_{ry}$ for some $r\in\mathcal{R}$ but it is the minimum among all feasible values of $M$. In this sense, \emph{the solution $x_{ry}^*$ guarantees fairness for all $r\in\mathcal{R}$}. However, sometimes the network connectivity, communication latency, and computation latency constraints may play a very strong role to produce an optimal solution $(x_{ry}^*,M^*)$ without maintaining fairness. Now, we analyze all possible network scenarios to show the validity of this proposition.\par
\textbf{Case 1:} \textit{Constraints (6), (9)-(12) are relaxed.}\par
This is the trivial case when there is full-mesh connectivity among the RUs and Edge/OLT-Clouds and no strict latency bound exists. Therefore, any RU can be connected to any Edge/OLT-Cloud and the optimal solution will be dictated by the objective (4) and constraints (5) and (13). Therefore, the optimum value of $M$ will be achieved only when all the RUs are connected to a single Edge/OLT-Cloud with $\min\{B_y^{UL}+B_y^{DL}\}$ and/or $\min\{G_y^{UL}+G_y^{DL}\}$. It is also obvious from (4) that the OPEX of each RU will be directly proportional to their resource demands $W_r^{UL}$, $W_r^{DL}$, $\Gamma_r^{UL}$, and $\Gamma_r^{DL}$, which ensure fairness.\par
\textbf{Case 2:} \textit{Constraint (6) is relaxed only.}\par
In this case, although there is full-mesh connectivity among the RUs and Edge/OLT-Clouds, strict communication and processing latency bounds are applied. Therefore, only a limited number of RUs can be connected to each Edge/OLT-Cloud to satisfy constraints (9)-(12). In this case, the best possible solution can be achieved only if all RUs can be connected to (one or multiple) Edge/OLT-Clouds with minimum total cost of resources. Without loss of generality, consider that there are total $|\mathcal{R}|$ number of RUs and 2 front/mid-haul links and Edge/OLT-Clouds. If the front/mid-haul and Edge/OLT-Cloud 1 have sufficient resources to host all $|\mathcal{R}|$ number of RUs, then there is no issue of fairness. However, if there are insufficient resources in Edge/OLT-Cloud 1 and some RUs need to be assigned to Edge/OLT-Cloud 2, then it is best to choose RUs with higher resource demands. By following this method, the fairness of the shared cost of each RU assigned to either of the Edge/OLT-Cloud can be guaranteed. If we assign RUs in any other randomized way, fairness cannot be guaranteed. For example, if we can assign only the RU with the lowest demand to Edge/OLT-Cloud 2 and the rest of the RUs to Edge/OLT-Cloud 1, then the optimal value $M^*$ remains the same (total cost of resources of front/mid-haul and Edge/OLT-Cloud 2), but the fairness is not maintained. Note that this argument remains valid even if we generalize our scenario with more than 2 Edge/OLT-Clouds.\par
\textbf{Case 3:} \textit{Constraint (9)-(12) are relaxed only.}\par
In this case, usually, there is partial-mesh connectivity among the RUs and Edge/OLT-Clouds but communication and processing latency bounds are relaxed. Therefore, in spite of sufficient resources being present, we cannot connect all the RUs to a single Edge/OLT-Cloud due to network connectivity constraints. This implies that the cost of shared resources of multiple RUs with the exact same resource demand may vary if they are connected to different Edge/OLT-Clouds by being forced by the network connectivity constraints. However, their shared costs will be proportionally fair in comparison with the other RUs connected to each Edge/OLT-Clouds. For example, let us consider two RUs with the exact same resource demands but allocated to two different Edge/OLT-Clouds and a different number of other RUs connected to these Edge/OLT-Clouds. Therefore, the OPEX values may be different for these RUs but their OPEX values will be fair in comparison to other RUs connected to their corresponding Edge/OLT-Clouds.\par
\textbf{Case 4:} \textit{All constraints (6), (9)-(12) are active.}\par
This is the most general case and characteristics of all the previous cases can be observed. Depending on the dataset, either constraint (6) or constraints (9)-(12) will dictate the optimal solution. Although sufficient resources may be present at some Edge/OLT-Cloud to serve a large number of RUs, constraint (6) might interfere. Therefore, in this case, also the cost of shared resources of multiple RUs with the exact same resource demand may vary if they are connected to different Edge/OLT-Clouds. However, the OPEX values of each RU will be proportionally fair to their resource demands in comparison with other RUs connected to each Edge/OLT-Clouds.\par 
In general, we observe by testing a number of diverse datasets that to obtain the best possible solution $M^*$ while maintaining the fairness of the OPEX values of the RUs, we must start to assign RUs to Edge/OLT-Clouds in the increasing order of their resource requirements.
\end{IEEEproof}

\section{Proof of Proposition 2} \label{sec2a}
\begin{IEEEproof}
We consider \emph{a resource allocation mechanism is allocatively efficient} if it optimizes the sum of the valuations of the agents for each given type profile of the agents \cite{Narahari}. Now, in our problem formulation $\mathcal{P}_2$, the total cost of front/mid-haul and Edge/OLT-Clouds are minimized. Although the binary variable $t_y$ indicates if an Edge/OLT-Cloud $y$ is activated or not, the constraint $t_y \leq x_{ry}, \forall r\in\mathcal{R}, y\in\mathcal{Y}$ makes this formulation equivalent to minimizing the sum of valuation functions $V_r(x_{ry})$ for all RUs. Therefore, our allocation rule $\hat{x}_{ry}^*$ derived as an optimal solution of $\mathcal{P}_2$ can be considered as allocatively efficient.
\end{IEEEproof}

\section{Proof of Theorem 1} \label{sec3a}
\begin{IEEEproof}
The proposed payment rule can be interpreted as the total value of all RUs other than $r$ under an efficient allocation when RU $r$ is absent in the system minus the total value of all RUs other than $r$ under an efficient allocation when RU $r$ is present in the system. Observe that each RU $r$ is connected to an Edge/OLT-Cloud $y$ based on the shared information $b_r = (\hat{W}_r^{UL}, \hat{W}_r^{DL}, \hat{\Gamma}_r^{UL}, \hat{\Gamma}_r^{DL})$ from all RUs.\par 
\textbf{Case 1:} If the shared information $b_r$ is \emph{higher than its actual requirement}, then there is a risk that it might remain unallocated when the network is in high-load condition. Then the valuation, payment, and utility of the RU $r$ are given by:
\begin{align}
    & V_r(\hat{x}_{ry}^*) = 0, \nonumber\\
    & P_r(\hat{x}_{ry}^*,\bm{b}) = 0, \nonumber\\
    & \mathcal{U}_r(\hat{x}_{ry}^*,\bm{b}) = 0. \nonumber
\end{align}
\par \textbf{Case 2:} If the shared information $b_r$ is \emph{lower than its actual requirement}, then the RU might get allocated to some Edge/OLT-Cloud, but due to insufficient resources in the RU's share, the front/mid-haul data generated per slot may fail to get delivered within the maximum one-way front/mid-haul latency requirements. In this case, the valuation is zero but the payment is non-zero, yielding a negative utility. Thus, the valuation, payment, and utility of the RU $r$ are:
\begin{align}
    & V_r(\hat{x}_{ry}^*) = 0, \nonumber\\
    & P_r(\hat{x}_{ry}^*,\bm{b}) = \sum_{j=1,j\neq r}^{|\mathcal{R}|} \left(\frac{C_\lambda B_y + C_P G_y}{\sum_{k\neq r} \hat{x}_{ky}^{-r*}}\right)\nonumber\\
    &\hspace{2cm}- \sum_{j=1,j\neq r}^{|\mathcal{R}|} \left(\frac{C_\lambda B_y + C_P G_y}{\sum_k \hat{x}_{ky}^*}\right), \nonumber\\
    & \mathcal{U}_r(\hat{x}_{ry}^*,\bm{b}) = 0 - P_r(\hat{x}_{ry}^*,\bm{b}) < 0. \nonumber
\end{align}
\par \textbf{Case 3:} There may arise network scenarios where an RU $r$ might get connected to some Edge/OLT-Cloud by sharing false information (both higher and lower) and its front/mid-haul data generated per slot is also successfully delivered within the maximum one-way front/mid-haul latency requirements. However, after the RU to Edge/OLT-Cloud allocation, when the payment amount is calculated for each RU $r$, only the total number of RUs connected to each Edge/OLT-Cloud $y$ is used and the revealed front/mid-haul datarate does not have any role to play. Thus, \emph{the utility of the RU with true revealed information as well as false revealed information remains the same}. In this case, the valuation, payment, and utility of the RU $r$ are given below:
\begin{align}
    & V_r(\hat{x}_{ry}^*) = \sum_y (C_{\lambda} B_y + C_P G_y)\hat{x}_{ry}^*, \nonumber\\
    & P_r(\hat{x}_{ry}^*,\bm{b}) = \sum_{j=1,j\neq r}^{|\mathcal{R}|} \left(\frac{C_\lambda B_y + C_P G_y}{\sum_{k\neq r} \hat{x}_{ky}^{-r*}}\right) \nonumber\\
    &\hspace{2cm}- \sum_{j=1,j\neq r}^{|\mathcal{R}|} \left(\frac{C_\lambda B_y + C_P G_y}{\sum_k \hat{x}_{ky}^*}\right), \nonumber\\
    & \mathcal{U}_r(\hat{x}_{ry}^*,\bm{b}) = V_r(\hat{x}_{ry}^*) - P_r(\hat{x}_{ry}^*,\bm{b}) \geq 0. \nonumber
\end{align}
\par Therefore, any RU $r$ can not gain any advantages in its utility with the payment rule by sharing false information with the NSP. This ensures that sharing truthful information with the NSP is a weakly dominant strategy for all the RUs.
\end{IEEEproof}
\ul{\textbf{Example 1:}} [One RU and One Edge/OLT-Cloud]\par
Let us consider the total x-haul throughput is $B_1$ and total Edge/OLT-Cloud resources is $G_1$. The RU 1 submits a message $b_1$ to the NSP. The results of Case 1 and Case 2 are straight-forward to show. If the scenario is like Case 3, then the RU 1 is connected to Edge/OLT-Cloud 1. Therefore the valuation of the RU is:
\begin{align}
    V_1 = (C_{\lambda} B_1 + C_P G_1). \nonumber
\end{align}
\par Now, the cost of resources for the RU 1 is $(C_{\lambda} B_1 + C_P G_1)$ as there is no other RU to share the cost of resources. However, if RU 1 was not present, then the entire block of resources would have remain unallocated with zero cost of resources. Hence, the actual payment made by RU 1 is given by:
\begin{align}
    P_1 &= \left(C_{\lambda} B_1 + C_P G_1\right) - 0 \nonumber\\
    &= \left(C_{\lambda} B_1 + C_P G_1\right). \nonumber
\end{align}
\par This shows that if an RU gets sufficient resources to transmit its radio data generated in each TTI, then its payment rule is independent of its shared private information. Therefore, sharing truthful information is a weakly dominant strategy in this scenario.\par
\begin{figure*}[!t]
  \centering
  \subfloat[]{%
    \includegraphics[width=0.333\textwidth]{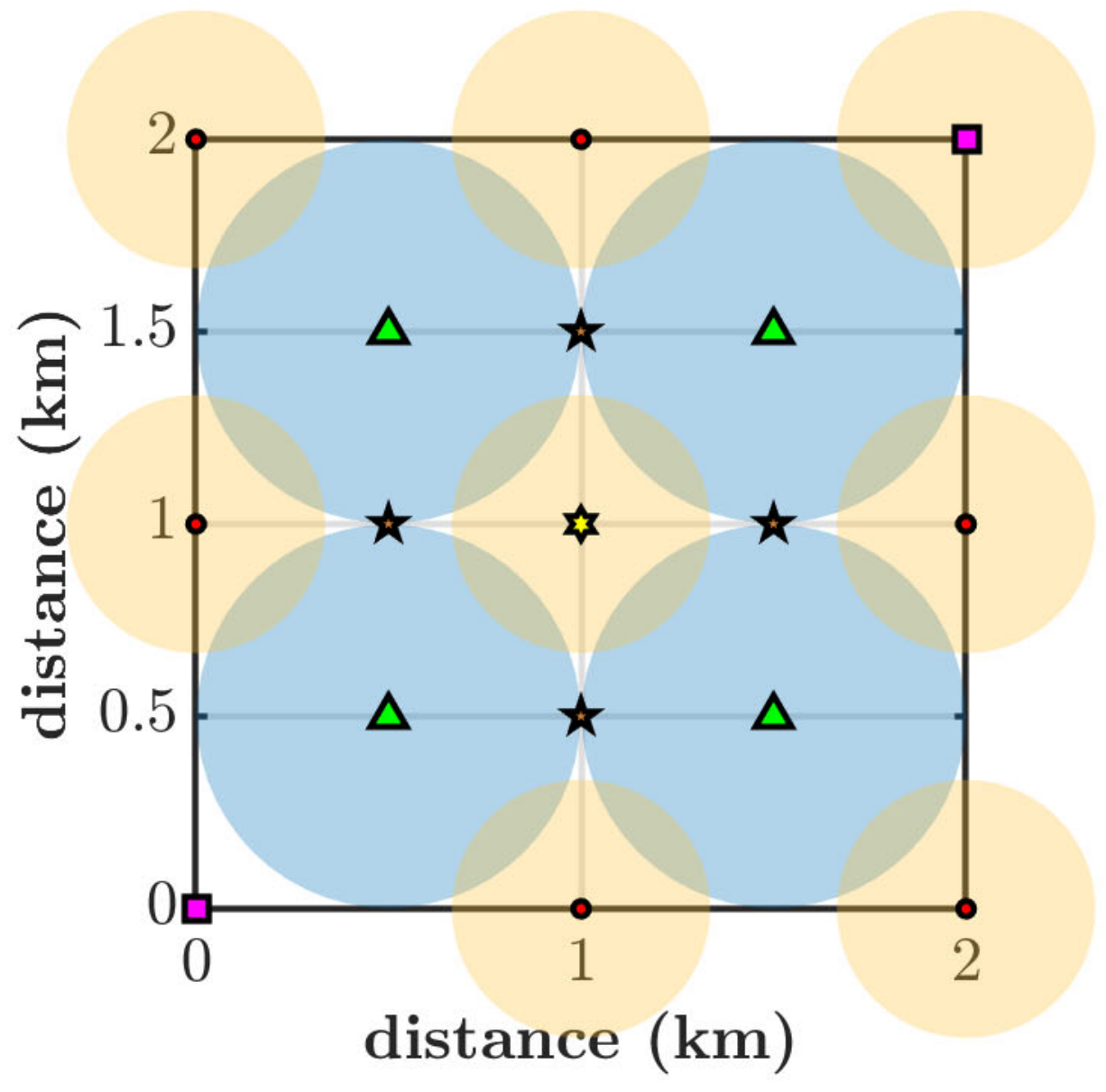}\label{cost4}%
  }
  \subfloat[]{%
    \includegraphics[width=0.333\textwidth]{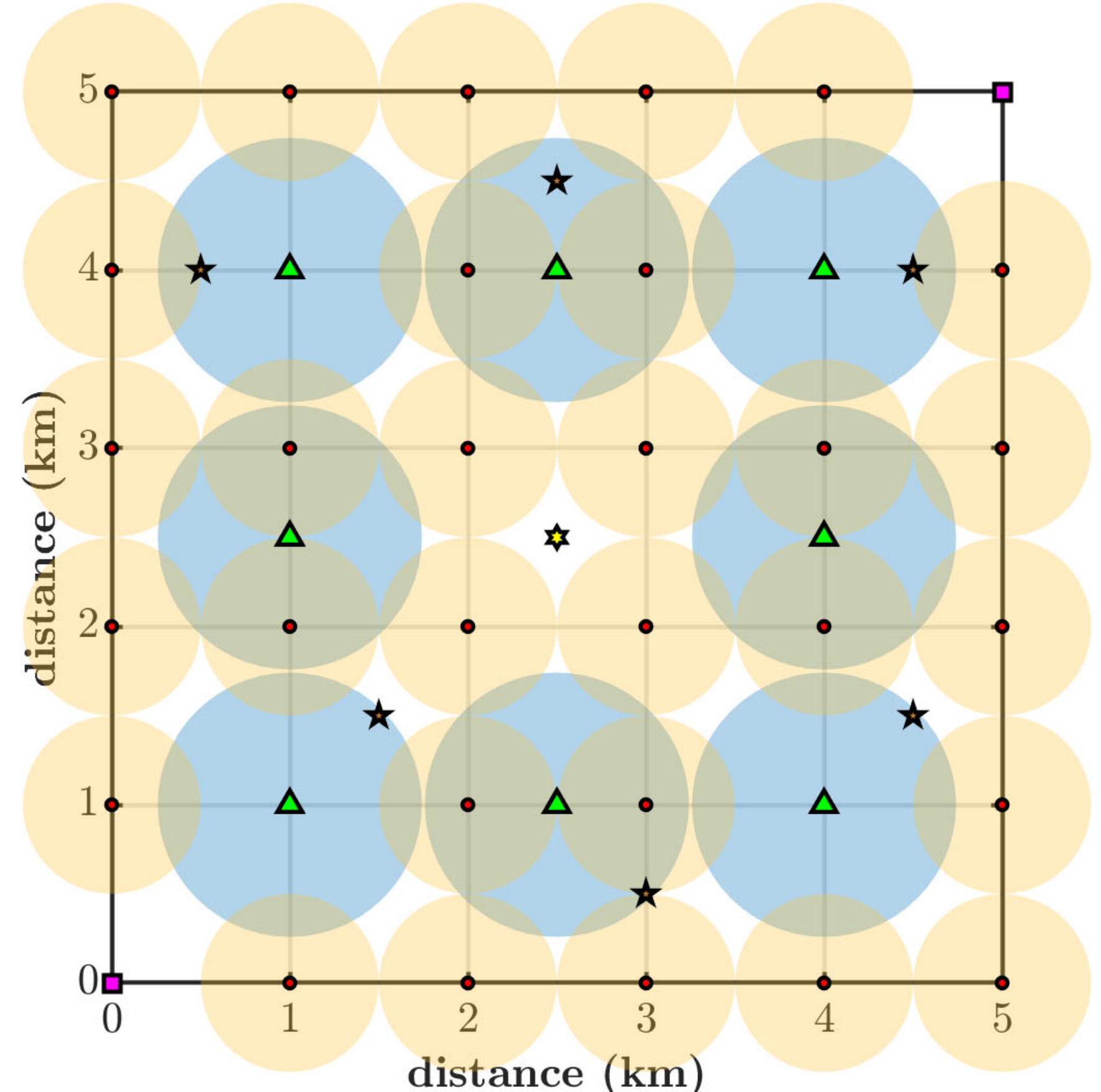}\label{cost5}%
  }
  \subfloat[]{%
    \includegraphics[width=0.333\textwidth]{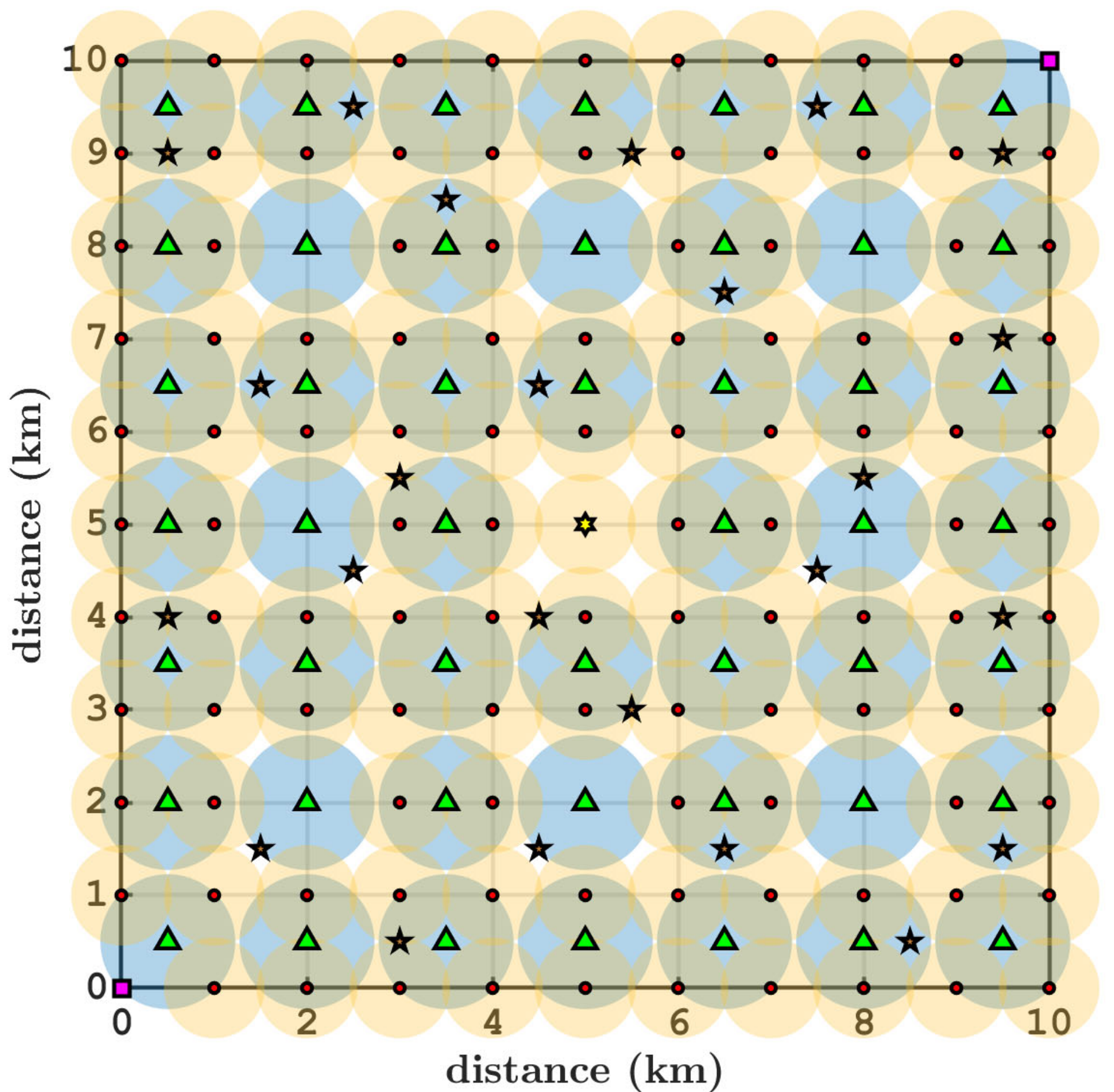}\label{cost6}%
  }

  \caption{Considered O-RAN deployment areas with two CO locations at diagonally opposite corner points with total dimension (a) 2 km $\times$ 2 km, (b) 5 km $\times$ 5 km, and (c) 10 km $\times$ 10 km (pink square: CO locations, orange pentagon: level-1 splitter, yellow hexagon: level-2 splitter, red circle: small-cell RU, green triangle: macro-cell RU).}
  \label{cost_compare}
\end{figure*}
\setlength{\textfloatsep}{1pt}
\ul{\textbf{Example 2:}} [Two RUs and One Edge/OLT-Cloud]\par
Let us consider the total x-haul throughput is $B_1$ and total Edge/OLT-Cloud resources is $G_1$. The RU 1 submits a message $b_1$ and the RU 2 submits a message $b_2$ where $b_1 \leq b_2$. The results of Case 1 and Case 2 can be shown in a straight-forward manner. If the scenario is like Case 3, then both the RUs are connected to Edge/OLT-Cloud 1. Therefore the valuations of both the RUs are:
\begin{align}
    V_1 = V_2 = (C_{\lambda} B_1 + C_P G_1). \nonumber
\end{align}
\par Now, the cost of resources for each of the RUs is $(C_{\lambda} B_1 + C_P G_1)/2$. However, if RU 1 was not present, then the entire block of resources would have been allocated to RU 2 with cost of resources = $(C_{\lambda} B_1 + C_P G_1)$. Therefore, the actual payment made by RU 1 is given by:
\begin{align}
    P_1 &= \left(C_{\lambda} B_1 + C_P G_1\right) - \left(\frac{C_{\lambda} B_1 + C_P G_1}{2}\right) \nonumber\\
    &= \left(\frac{C_{\lambda} B_1 + C_P G_1}{2}\right). \nonumber
\end{align}
\par A similar analysis is applicable for RU 2. This shows that if an RU gets sufficient resources to transmit its radio data generated in each TTI, then its payment rule is independent of its shared private information. Therefore, sharing truthful information is a weakly dominant strategy in this scenario.\par
\ul{\textbf{Example 3:}} [Three RUs and Two Edge/OLT-Clouds]\par
Let us consider the total x-haul throughput are $B_1$, $B_2$ and total Edge/OLT-Cloud resources are $G_1$, $G_2$. The RUs submit messages $(b_1, b_2,b_3)$ to the NSP where $b_1 \leq b_2 \leq b_3$. The results of Case 1 and Case 2 are straight-forward to show. If the scenario is like Case 3 and we assume that \emph{RU 1 and RU 2 are allocated to Edge-Cloud 1 whereas RU 3 is allocated to Edge-Cloud 2}, then the valuations of all the RUs are:
\begin{align}
    & V_1 = V_2 = (C_{\lambda} B_1 + C_P G_1), \nonumber\\
    & V_3 = (C_{\lambda} B_2 + C_P G_2). \nonumber
\end{align}
\par Now, the cost of shared resources for each of the RUs are given as follows:
\begin{align}
    & \tilde{\mathcal{C}}_1 = \tilde{\mathcal{C}}_2 = (C_{\lambda} B_1 + C_P G_1)/2, \nonumber\\
    & \tilde{\mathcal{C}}_3 = (C_{\lambda} B_2 + C_P G_2). \nonumber
\end{align}
\par However, if RU 1 was not present, then two cases can happen. Either RU 3 is allocated to Edge/OLT-Cloud 1 along with RU 2 while Edge/OLT-Cloud 2 remains inactive, or RU 2 is allocated to Edge/OLT-Cloud 1 while RU 2 is allocated to Edge/OLT-Cloud 2. Nonetheless, in both the cases, the total cost of resources to be paid by RUs allocated to Edge/OLT-Clouds 1 is $(C_{\lambda} B_1 + C_P G_1)$. Therefore, the actual payment made by RU 1 is given by:
\begin{align}
    P_1 &= \left(C_{\lambda} B_1 + C_P G_1\right) - \left(\frac{C_{\lambda} B_1 + C_P G_1}{2}\right) \nonumber\\
    &= \left(\frac{C_{\lambda} B_1 + C_P G_1}{2}\right). \nonumber
\end{align}
\par We can analyze the payment rules for RU 2 and RU 3 by following the previous examples. Furthermore, \emph{if all the three RUs were allocated to Edge/OLT-Cloud 1} in the first place, then we can show through a similar analysis as is Example 2 that the payment made by RU 1 is given by:
\begin{align}
    P_1 &= \left[\left(\frac{C_{\lambda} B_1 + C_P G_1}{2}\right)+\left(\frac{C_{\lambda} B_1 + C_P G_1}{2}\right)\right] \nonumber\\
    & \quad\quad\quad -\left[\left(\frac{C_{\lambda} B_1 + C_P G_1}{3}\right)+\left(\frac{C_{\lambda} B_1 + C_P G_1}{3}\right)\right] \nonumber\\
    &= \left(C_{\lambda} B_1 + C_P G_1\right) - \frac{2}{3}\left(C_{\lambda} B_1 + C_P G_1\right) \nonumber\\
    &= \left(\frac{C_{\lambda} B_1 + C_P G_1}{3}\right). \nonumber
\end{align}
\par This clearly shows that if an RU gets sufficient resources to transmit its radio data generated in each TTI, then its payment rule is independent of its shared private information. Therefore, sharing truthful information is a weakly dominant strategy in this scenario.

\section{Considered O-RAN Deployment Areas} \label{sec4a}
In this section, we graphically show some O-RAN deployment areas that are considered for the numerical evaluation of our proposed frameworks. Firstly, Fig. \ref{cost4} shows an O-RAN deployment area of dimension  2 km $\times$ 2 km, secondly, Fig. \ref{cost5} shows an O-RAN deployment area of dimension  5 km $\times$ 5 km, and thirdly, Fig. \ref{cost6} shows an O-RAN deployment area of dimension 10 km $\times$ 10 km. In all these figures, each macro-cell RU has an approximate coverage area of 1.0 km (shown using light-blue circles) and each small-cell RU has an approximate coverage area of 0.5 km (shown using light-yellow circles). Also, a few symbols are used to denote key elements, i.e., pink square: CO locations, orange pentagon: level-1 splitter, yellow hexagon: level-2 splitter, red circle: small-cell RU, and green triangle: macro-cell RU. The actual fiber connections are not shown to avoid the figures from being over-crowded and stingy. The coordinates of the two CO locations are given by $(0,0)$ and $(D,D)$ where $D\in\{2,5,10\}$. A group of TWDM-PONs branch out from each of the CO locations that are used to connect the RUs, e.g., the RUs located below the primary diagonal are all connected to the CO at $(0,0)$, but the the RUs located above the primary diagonal are all connected to the CO at $(D,D)$. Although a rectangular grid-like deployment of RUs are considered for our proposed framework evaluation, any practical RU deployment data can be used as an input to this framework.


\bibliographystyle{IEEEtran}
\bibliography{IEEEabrv,references}

\begin{thebibliography}{10}
\providecommand{\url}[1]{#1}
\csname url@samestyle\endcsname
\providecommand{\newblock}{\relax}
\providecommand{\bibinfo}[2]{#2}
\providecommand{\BIBentrySTDinterwordspacing}{\spaceskip=0pt\relax}
\providecommand{\BIBentryALTinterwordstretchfactor}{4}
\providecommand{\BIBentryALTinterwordspacing}{\spaceskip=\fontdimen2\font plus
\BIBentryALTinterwordstretchfactor\fontdimen3\font minus
  \fontdimen4\font\relax}
\providecommand{\BIBforeignlanguage}[2]{{%
\expandafter\ifx\csname l@#1\endcsname\relax
\typeout{** WARNING: IEEEtran.bst: No hyphenation pattern has been}%
\typeout{** loaded for the language `#1'. Using the pattern for}%
\typeout{** the default language instead.}%
\else
\language=\csname l@#1\endcsname
\fi
#2}}
\providecommand{\BIBdecl}{\relax}
\BIBdecl

\bibitem{6G_vision}
W.~{Saad}, M.~{Bennis}, and M.~{Chen}, ``{A Vision of 6G Wireless Systems:
  Applications, Trends, Technologies, and Open Research Problems},'' \emph{IEEE
  Network}, vol.~34, no.~3, pp. 134--142, 2020.

\bibitem{oran}
``{O-RAN: Towards an Open and Smart RAN},'' O-RAN Alliance, Tech. Rep., Oct
  2018.

\bibitem{ngfi2}
C.-L. I \emph{et~al.}, ``{RAN Revolution With NGFI (xhaul) for 5G},'' \emph{J.
  Lightw. Technol.}, vol.~36, no.~2, pp. 541--550, 2018.

\bibitem{aceg1}
L.~Peterson \emph{et~al.}, ``{Democratizing the Network Edge},'' \emph{ACM
  SIGCOMM Comp. Commun. Rev.}, vol.~49, no.~2, p. 31–36, May 2019.

\bibitem{smsng}
\BIBentryALTinterwordspacing
``{The Open Road to 5G},'' Samsung Networks, Tech. Rep., July 2019. [Online].
  Available:
  \url{https://image-us.samsung.com/SamsungUS/samsungbusiness/pdfs/Open-RAN-The-Open-Road-to-5G.pdf}
\BIBentrySTDinterwordspacing

\bibitem{tech_eco}
B.~Naudts \emph{et~al.}, ``{Techno-economic Analysis of Software Defined
  Networking as Architecture for the Virtualization of a Mobile Network},'' in
  \emph{2012 European Workshop Softw. Defined Netw.}, 2012, pp. 67--72.

\bibitem{pon_oran}
T.~Pfeiffer, P.~Dom, S.~Bidkar, F.~Fredricx, K.~Christodoulopoulos, and
  R.~Bonk, ``{PON going beyond FTTH [Invited Tutorial]},'' \emph{J. Opt.
  Commun. Netw.}, vol.~14, no.~1, pp. A31--A40, 2022.

\bibitem{sandip}
S.~Das \emph{et~al.}, ``{Virtualized EAST–WEST PON Architecture Supporting
  Low-latency Communication for Mobile Functional Split based on Multiaccess
  Edge Computing},'' \emph{IEEE/OSA J. Opt. Commun. Netw.}, vol.~12, no.~10,
  pp. D109--D119, 2020.

\bibitem{fair}
L.~Massoulie and J.~Roberts, ``{Bandwidth Sharing: Objectives and
  Algorithms},'' \emph{IEEE/ACM Trans. Netw.}, vol.~10, no.~3, pp. 320--328,
  2002.

\bibitem{nyto_auct}
D.~Niyato, N.~C. Luong, P.~Wang, and Z.~Han, ``{Second-Price Sealed-Bid
  Auction},'' in \emph{Auction Theory for Computer Networks}.\hskip 1em plus
  0.5em minus 0.4em\relax Cambridge University Press, 2020, ch.~6, pp.
  119--157.

\bibitem{res_alloc_bk2}
L.~Tan, \emph{{Resource Allocation and Performance Optimization in
  Communication Networks and the Internet}}.\hskip 1em plus 0.5em minus
  0.4em\relax USA: CRC Press, Inc., 2017.

\bibitem{oran_app}
D.~Wypiór, M.~Klinkowski, and I.~Michalski, ``{Open RAN—Radio Access Network
  Evolution, Benefits and Market Trends},'' \emph{Applied Sciences}, vol.~12,
  no.~1, 2022.

\bibitem{yao}
J.~Yao and N.~Ansari, ``{QoS-Aware Joint BBU-RRH Mapping and User Association
  in Cloud-RANs},'' \emph{IEEE Trans. Green Commun. Netw.}, vol.~2, no.~4, pp.
  881--889, 2018.

\bibitem{cran_mgmt}
M.~Barahman \emph{et~al.}, ``{A Fair Computational Resource Management Strategy
  in C-RAN},'' in \emph{2018 International Conference on Broadband
  Communications for Next Generation Networks and Multimedia Applications
  (CoBCom)}, 2018, pp. 1--6.

\bibitem{cran_opt1}
N.~Mharsi and M.~Hadji, ``{Joint Optimization of Communication Latency and
  Resource Allocation in Cloud Radio Access Networks},'' in \emph{2018
  International Conf. on Smart Commun. in Netw. Technol. (SaCoNeT)}, 2018, pp.
  13--18.

\bibitem{cran_opt2}
M.~Peng, Y.~Yu, H.~Xiang, and H.~V. Poor, ``{Energy-Efficient Resource
  Allocation Optimization for Multimedia Heterogeneous Cloud Radio Access
  Networks},'' \emph{IEEE Transactions on Multimedia}, vol.~18, no.~5, pp.
  879--892, 2016.

\bibitem{cran_ra}
M.~Y. Lyazidi, N.~Aitsaadi, and R.~Langar, ``{Dynamic resource allocation for
  Cloud-RAN in LTE with real-time BBU/RRH assignment},'' in \emph{2016 IEEE
  Int. Conf. Commun. (ICC)}, 2016, pp. 1--6.

\bibitem{rrh-bbu1}
S.~Guo, D.~Zeng, L.~Gu, and J.~Luo, ``{When Green Energy Meets Cloud Radio
  Access Network: Joint Optimization Towards Brown Energy Minimization},''
  \emph{Mob. Netw. Appl.}, vol.~24, no.~3, p. 962–970, Jun. 2019.

\bibitem{rrh-bbu4}
K.~Boulos, M.~El~Helou, and S.~Lahoud, ``{RRH Clustering in Cloud Radio Access
  Networks},'' in \emph{2015 International Conference on Applied Research in
  Computer Science and Engineering (ICAR)}, 2015, pp. 1--6.

\bibitem{rru-bbu5}
H.~M. Soliman and A.~Leon-Garcia, ``{QoS-aware Joint RRH Activation and
  Clustering in Cloud-RANs},'' in \emph{2016 IEEE Wireless Commun. Netw. Conf.
  (WCNC)}, 2016, pp. 1--6.

\bibitem{tenant1}
D.~Liang, R.~Gu, Q.~Guo, and Y.~Ji, ``{Demonstration of Multi-Vendor
  Multi-Standard PON Networks for Network Slicing in 5G-oriented Mobile
  Network},'' in \emph{2017 Asia Commun. Photon. Conf. (ACP)}, 2017.

\bibitem{oran_srvy}
L.~Bonati, M.~Polese, S.~D’Oro, S.~Basagni, and T.~Melodia, ``{Open,
  Programmable, and Virtualized 5G Networks: State-of-the-Art and the Road
  Ahead},'' \emph{Comput. Netw.}, vol. 182, p. 107516, 2020.

\bibitem{oran_snsr}
M.~Dryjański, u.~Kułacz, and A.~Kliks, ``{Toward Modular and Flexible Open
  RAN Implementations in 6G Networks: Traffic Steering Use Case and O-RAN
  xApps},'' \emph{Sensors}, vol.~21, no.~24, 2021.

\bibitem{oran_sm_tnsm}
S.~Mondal and M.~Ruffini, ``{Optical Front/Mid-haul with Open Access-Edge
  Server Deployment Framework for Sliced O-RAN},'' \emph{IEEE Transactions on
  Network and Service Management}, pp. 1--18, 2022.

\bibitem{oran_tnsm}
M.~K. Motalleb, V.~Shah-Mansouri, S.~Parsaeefard, and O.~L.~A. López,
  ``{Resource Allocation in an Open RAN System using Network Slicing},''
  \emph{IEEE Transactions on Network and Service Management}, pp. 1--1, 2022.

\bibitem{oran_ra}
X.~Wang, J.~D. Thomas, R.~J. Piechocki, S.~Kapoor, R.~Santos-Rodríguez, and
  A.~Parekh, ``{Self-play learning strategies for resource assignment in
  Open-RAN networks},'' \emph{Computer Networks}, vol. 206, p. 108682, 2022.

\bibitem{oran_team}
H.~Zhang, H.~Zhou, and M.~Erol-Kantarci, ``{Team Learning-Based Resource
  Allocation for Open Radio Access Network (O-RAN)},'' in \emph{ICC 2022 - IEEE
  International Conference on Communications}, 2022, pp. 4938--4943.

\bibitem{mlt_op_ran}
M.~Kassis \emph{et~al.}, ``{Flexible Multi-Operator RAN Sharing:
  Experimentation and Validation Using Open Source 4G/5G Prototype},'' in
  \emph{2021 Joint European Conf. Netw. and Commun. \& 6G Summit (EuCNC/6G
  Summit)}, 2021, pp. 205--210.

\bibitem{WirelessMoves}
M.~Sauter, ``{Long Term Evolution (LTE) and LTE-Advanced Pro},'' in \emph{From
  GSM to LTE‐Advanced Pro and 5G}.\hskip 1em plus 0.5em minus 0.4em\relax
  John Wiley \& Sons Ltd., 2017, ch.~4, pp. 211--333.

\bibitem{tenant2}
R.~Vilalta \emph{et~al.}, ``{Network Virtualization Controller for Abstraction
  and Control of OpenFlow-enabled Multi-tenant Multi-technology Transport
  Networks},'' in \emph{Opt. Netw. Commun. Conf. Exhibit. (OFC)}, 2015.

\bibitem{sm_mnmx}
S.~Mondal and M.~Ruffini, ``{A Min-Max Fair Resource Allocation Framework for
  Optical x-haul and DU/CU in Multi-tenant O-RANs},'' in \emph{2022 IEEE Int.
  Conf. Commun. (ICC)}, May 2022, pp. 452--457.

\bibitem{itu_5g}
\BIBentryALTinterwordspacing
``{5G Wireless Fronthaul Requirements in a Passive Optical Network Context},''
  Telecommunication Standardization Sector of ITU (ITU-T), Tech. Rep., Sep
  2020. [Online]. Available:
  \url{https://www.itu.int/rec/dologin_pub.asp?lang=e&id=T-REC-G.Sup66-202009-I!!PDF-E&type=items}
\BIBentrySTDinterwordspacing

\bibitem{marco1}
M.~Ruffini \emph{et~al.}, ``{Virtual DBA: Virtualizing Passive Optical Networks
  to Enable Multi-service Operation in True Multi-tenant Environments},''
  \emph{IEEE/OSA J. Opt. Commun. Netw.}, vol.~12, no.~4, pp. B63--B73, 2020.

\bibitem{marco2}
F.~Slyne, S.~Zeb, and M.~Ruffini, ``{Stateful DBA Hypervisor Supporting SLAs
  with Low Latency and High Availability in Shared PON},'' in \emph{2021 Opt.
  Netw. Commun. Conf. Exhibit. (OFC)}, 2021, pp. 1--3.

\bibitem{colo}
R.~{Beraldi}, A.~{Mtibaa}, and H.~{Alnuweiri}, ``{Cooperative Load Balancing
  Scheme for Edge Computing Resources},'' in \emph{2017 Second International
  Conference on Fog and Mobile Edge Computing (FMEC)}, May 2017, pp. 94--100.

\bibitem{SCF}
\BIBentryALTinterwordspacing
``{Small Cell Virtualization Functional Splits and Use Cases},'' 2015.
  [Online]. Available:
  \url{https://scf.io/en/documents/159_-_Small_Cell_Virtualization_Functional_Splits_and_Use_Cases.php}
\BIBentrySTDinterwordspacing

\bibitem{38.801}
\BIBentryALTinterwordspacing
``{Study on new radio access technology: Radio access architecture and
  interfaces},'' 2017, accessed: 2022-07-02. [Online]. Available:
  \url{https://www.3gpp.org/DynaReport/38801.htm}
\BIBentrySTDinterwordspacing

\bibitem{5g_fh_bw3}
T.~{Tashiro}, S.~{Kuwano}, J.~{Terada}, T.~{Kawamura}, N.~{Tanaka},
  S.~{Shigematsu}, and N.~{Yoshimoto}, ``{A Novel DBA Scheme for TDM-PON based
  Mobile Fronthaul},'' in \emph{2014 Opt. Netw. Commun. Conf. Exhibit. (OFC)},
  2014, pp. 1--3.

\bibitem{mgain}
M.~Shehata \emph{et~al.}, ``{Multiplexing Gain and Processing Savings of 5G
  Radio-Access-Network Functional Splits},'' \emph{IEEE Trans. Green Commun.
  Netw.}, vol.~2, no.~4, pp. 982--991, 2018.

\bibitem{bbu_lat}
S.~Khatibi, K.~Shah, and M.~Roshdi, ``{Modelling of Computational Resources for
  5G RAN},'' in \emph{2018 European Conference on Networks and Communications
  (EuCNC)}, 2018, pp. 1--5.

\bibitem{moo}
E.~K.~P. Chong and S.~H. Żak, ``Multiobjective optimization,'' in \emph{An
  Introduction to Optimization}.\hskip 1em plus 0.5em minus 0.4em\relax John
  Wiley \& Sons, Ltd, 2008, ch.~23, pp. 541--562.

\bibitem{mult_auc}
L.~M. Ausubel and P.~Cramton, ``{Auctioning Many Divisible Goods},''
  \emph{Journal of the European Economics Association}, vol.~2, pp. 480--493,
  2004.

\bibitem{Narahari}
Y.~Narahari, ``{Vickrey-Clarke-Groves (VCG) Mechanisms},'' in \emph{{Game
  Theory and Mechanism Design}}.\hskip 1em plus 0.5em minus 0.4em\relax World
  Scientific Publishing Company Pvt. Ltd., 2014, ch.~18, pp. 267--284.

\bibitem{Sutton}
R.~S. Sutton and A.~G. Barto, \emph{{Introduction to Reinforcement Learning}},
  1st~ed.\hskip 1em plus 0.5em minus 0.4em\relax Cambridge, MA, USA: MIT Press,
  1998.

\bibitem{delay_calc}
D.~H. Hailu \emph{et~al.}, ``{Mobile fronthaul transport options in C-RAN and
  emerging research directions: A comprehensive study},'' \emph{Optical
  Switching and Networking}, vol.~30, pp. 40--52, 2018.

\bibitem{5g_fh_bw}
Y.~{Nakayama} \emph{et~al.}, ``{Efficient DWBA Algorithm for TWDM-PON with
  Mobile Fronthaul in 5G Networks},'' in \emph{2017 IEEE Global Commun. Conf.
  (GLOBECOM)}, 2017, pp. 1--6.

\bibitem{fib_rent}
``{Preliminary Fiber Network Design and Business Plan Framework},'' Columbia
  Telecommunications Corporation, Tech. Rep., Jun 2005.

\bibitem{F_split}
Y.~Xiao \emph{et~al.}, ``{Can Fine-Grained Functional Split Benefit to the
  Converged Optical-Wireless Access Networks in 5G and Beyond?}'' \emph{IEEE
  Trans. Netw. Service Manag.}, vol.~17, no.~3, pp. 1774--1787, 2020.

\end{thebibliography}
\vspace{-5ex}
\begin{IEEEbiography}[{\includegraphics[width=1in,height=1.25in,clip,keepaspectratio]{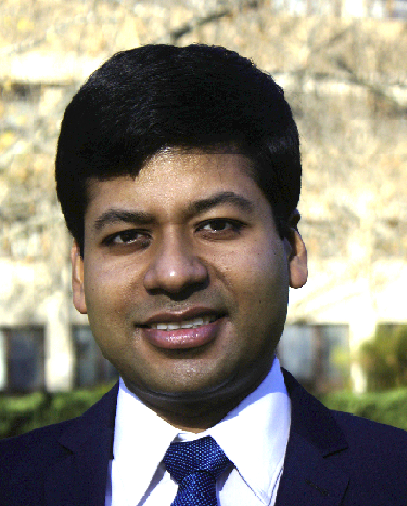}}]{Sourav Mondal}
(GS'16–M'21) received PhD from the Department of Electrical and Electronic Engineering of the University of Melbourne in 2020. He received his M.Tech in Telecommunication Systems Engineering from the Department of Electronics and Electrical Communication Engineering, Indian Institute of Technology Kharagpur and B.Tech in Electronics and Communication Engineering from Kalyani Govt. Engineering College, affiliated to West Bengal University of Technology in 2014 and 2012, respectively. He was employed as an Engineer in Qualcomm India Pvt. Ltd. from 2014 to 2016. Currently, he is working as an EDGE/Marie Skłodowska-Curie post-doctoral fellow at CONNECT Centre for Future Networks and Communication in Trinity College Dublin, Ireland.
\end{IEEEbiography} 
\begin{IEEEbiography}[{\includegraphics[width=1in,height=1.25in,clip,keepaspectratio]{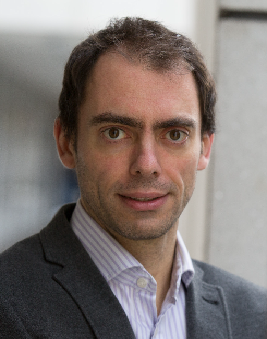}}]{Marco Ruffini} 
received the M.Eng. degree in telecommunications from the Polytechnic University of Marche, Italy, in 2002, and the Ph.D. degree Trinity College Dublin (TCD) in 2007, where he joined Trinity College Dublin in 2005, after working as a Research Scientist with Philips, Germany. He is Associate Professor and Fellow of Trinity College and he is Principal Investigator of both the IPIC Photonics Integration Centre and the CONNECT Telecommunications Research Centre. He is currently involved in several Science Foundation Ireland and H2020 projects, including a new research infrastructure to build a beyond 5G testbed in Dublin. Prof. Ruffini leads the Optical Network Architecture Group, TCD and has authored over 150 international publications, over ten patents and contributed to standards at the broadband forum. He has raised research funding in excess of \euro 7M. His main research is in the area of 5G optical networks, where he carries out pioneering work on the convergence of fixed-mobile and access-metro networks, and on the virtualization of next generation networks, and has been invited to share his vision through several keynote and talks at major international conferences across the world. He leads the new SFI funded Ireland’s Open Networking testbed infrastructure.
\end{IEEEbiography}

\vfill


\end{document}